\documentclass[twocolumn, longauthor]{aastex61}
\pdfoutput=1 %for arXiv submission
\usepackage{amsmath,amstext}
\usepackage[T1]{fontenc}
\usepackage{apjfonts} 
\usepackage[figure,figure*]{hypcap}

\usepackage{graphicx}	
\usepackage{amssymb}	
\usepackage{booktabs}
\usepackage{chemfig}

\accepted{March 30, 2018}
\submitjournal{ApJ}

\shorttitle{Water Prominence in Cool Giant Planets}
\shortauthors{MacDonald et al.}

\begin{document}

\title{Exploring H$_2$O Prominence in Reflection Spectra of Cool Giant Planets}

\correspondingauthor{Ryan J. MacDonald}
\email{r.macdonald@ast.cam.ac.uk}

\author[0000-0003-4816-3469]{Ryan J. MacDonald}
\affiliation{Institute of Astronomy \\
University of Cambridge \\
Cambridge, CB3 0HA, UK}

\author[0000-0002-5251-2943]{Mark S. Marley}
\affiliation{NASA Ames Research Center \\
Moffett Field, CA 94035, USA}

\author[0000-0002-9843-4354]{Jonathan J. Fortney}
\affil{University of California \\
1156 High Street \\
Santa Cruz, CA 95064, USA}

\author[0000-0002-8507-1304]{Nikole K. Lewis}
\affil{Space Telescope Science Institute \\
3700 San Martin Drive \\
Baltimore, MD 21218, USA}

\begin{abstract}

The H$_2$O abundance of a planetary atmosphere is a powerful indicator of formation conditions. Inferring H$_2$O in the solar system giant planets is challenging, due to condensation depleting the upper atmosphere of water vapour. Substantially warmer hot Jupiter exoplanets readily allow detections of H$_2$O via transmission spectroscopy, but such signatures are often diminished by the presence of clouds of other species. In contrast, highly scattering H$_2$O clouds can brighten planets in reflected light, enhancing molecular signatures. Here, we present an extensive parameter space survey of the prominence of H$_2$O absorption features in reflection spectra of cool ($T_{\mathrm{eff}} < 400$K) giant exoplanetary atmospheres. The impact of effective temperature, gravity, metallicity, and sedimentation efficiency is explored. We find prominent H$_2$O features around $0.94 \, \micron$, $0.83 \, \micron$, and across a wide spectral region from $0.4 - 0.73 \, \micron$. The 0.94 $\micron$ feature is only detectable where high-altitude water clouds brighten the planet: $T_{\mathrm{eff}} \sim 150$ K, $g \gtrsim$ 20 ms$^{-2}$, $f_{\mathrm{sed}} \gtrsim 3$, $m \lesssim 10 \times$ solar. In contrast, planets with $g \lesssim$ 20 ms$^{-2}$ and $T_{\mathrm{eff}} \gtrsim 180$ K display substantially prominent H$_2$O features embedded in the Rayleigh scattering slope from $0.4 - 0.73 \, \micron$ over a wide parameter space. High $f_{\mathrm{sed}}$ enhances H$_2$O features around $0.94 \, \micron$, and enables these features to be detected at lower temperatures. High $m$ results in dampened H$_2$O absorption features, due to H$_2$O vapour condensing to form bright optically thick clouds that dominate the continuum. We verify these trends via self-consistent modelling of the low gravity exoplanet HD 192310c, revealing that its reflection spectrum is expected to be dominated by H$_2$O absorption from $0.4 - 0.73 \, \micron$ for $m \lesssim 10 \times$ solar. Our results demonstrate that H$_2$O is manifestly detectable in reflected light spectra of cool giant planets only marginally warmer than Jupiter, providing an avenue to directly constrain the C/O and O/H ratios of a hitherto unexplored population of exoplanetary atmospheres.

\end{abstract}

\keywords{planets and satellites: atmospheres}

\section{Introduction} \label{sec:intro}

Spectroscopy of exoplanetary atmospheres has opened a new frontier in planetary science. The combined efforts of space-based and ground-based telescopes have resulted in conclusive detections of H$_2$O, CH$_4$, CO, TiO, Na, and K \citep[e.g][]{Deming2013,Macintosh2015,Snellen2010,Sedaghati2017,Snellen2008,Wilson2015}, along with evidence of species such as VO, HCN, and NH$_3$ \citep[e.g][]{Evans2017,MacDonald2017b}. Extracting abundances of these trace species from spectroscopic observations has been made possible by atmospheric retrieval techniques for exoplanets \citep[e.g][]{Madhusudhan2009,Line2013,Lupu2016}. The abundance of H$_2$O, in particular, has received intense attention as a potential diagnostic of planetary formation mechanisms \citep{Oberg2011,VanDishoeck2014}.

Comparative studies of H$_2$O in giant planet atmospheres has been pioneered by analyses of hot Jupiters \citep{Madhusudhan2014}. The high temperatures of these worlds ($\sim 1000-3000$ K) renders H$_2$O into vapour form throughout the atmosphere, resulting in strong infrared absorption features routinely observed in transmission and emission spectra \citep{Kreidberg2014b,Sing2016}. More recently, detections of H$_2$O have been extended to cooler exo-Neptunes ($\sim 1000$ K) \citep{Fraine2014,Wakeford2017} and thermally bright directly imaged planets in the outermost regions of young stars ($\sim 1000$ K) \citep{Konopacky2013,Barman2015}. At cooler temperatures, such as those of the solar system giant planets ($\sim 100$ K), H$_2$O condenses deep in the atmosphere, depleting their upper atmospheres of detectable quantities of H$_2$O vapour \citep{Atreya1999}.

Most H$_2$O absorption features observed via transmission spectroscopy are lower-amplitude than expected of a solar-composition cloud-free atmosphere \citep{Sing2016}. Such diminished features could be caused by a low atmospheric O abundance \citep{Madhusudhan2014}, or by obscuration due to high altitude clouds or hazes \citep{Deming2013}. Due to the slant geometry during primary transit, even a relatively vertically-thin cloud deck can result in large optical depths \citep{Fortney2005}. In extreme cases, such obscuration can result in characteristically flat transmission spectra devoid of spectral features \citep{Ehrenreich2014,Knutson2014,Kreidberg2014}. Where low-amplitude H$_2$O features are detected, the possibility of clouds can incur additional degeneracies when retrieving transmission spectra, resulting in weaker abundance constraints \citep{Benneke2015}.

Reflection spectroscopy offers an alternative route to characterise exoplanetary atmospheres. Contrasting with transmission spectra, early theoretical studies predicted that clouds can enhance molecular absorption features observed in reflected light \citep{Marley1999,Sudarsky2000,Burrows2004,Sudarsky2005}. This enhancement is caused by backscattering of photons from highly reflective clouds, enabling light that would otherwise have been consumed in the deep atmosphere to instead reach the observer. The resulting effect is an overall brightening of the planet and the accrual of absorption features from backscattered photons traversing the observable atmosphere.

Reflection spectra of Jupiter do not reveal signatures of H$_2$O \citep{Karkoschka1998}. At such low temperatures ($T_{\mathrm{eff}} \sim 100$ K) both NH$_3$ and H$_2$O clouds form, with the H$_2$O clouds sufficiently deep ($\sim$ 5 bar) as to render the observable upper atmosphere depleted of H$_2$O vapour. In-situ constraints on the Jovian H$_2$O abundance from the Galileo entry probe reported a value of 0.3$\times$ solar, contrasting with 2-3$\times$ solar measured for C, N, S, Ar, Kr, and Xe \citep{Wong2004}. This is surprising, as core accretion formation models assuming a solar-composition nebula predict an enhanced H$_2$O abundance of 3-7$\times$ solar \citep{Mousis2012}. Resolving the question of the global Jovian O abundance is one of the key science goals of NASA's Juno mission \citep{Bolton2010}.

Cool giant planets marginally warmer than Jupiter are expected to display signatures of H$_2$O in reflected light. Early studies noted than exoplanets with $T_{\mathrm{eff}} \sim 200$ K posses reflection spectra showing H$_2$O absorption at 0.94 $\micron$ \citep{Marley1999,Sudarsky2000,Burrows2004}. More recent studies confirmed that H$_2$O signatures are apparent for semi-major axes $<$ 5 AU \citep{Cahoy2010} and for thin, sufficiently deep, water clouds \citep{Morley2015}. \citet{Burrows2014} additionally noted that such H$_2$O signatures could serve as a diagnostic for metallicity. However, there has yet to be a thorough investigation of the prominence of H$_2$O signatures in cool giant planet reflection spectra across a wide range of planetary properties. Identifying regions of parameter space with detectable H$_2$O informs future direct imaging missions, such as WFIRST \citep{Spergel2015}, LUVOIR \citep{Bolcar2016}, HabEx \citep{Mennesson2016}, and similar concepts. Direct imaging spectroscopy in this manner will constrain the O abundances of an unexplored population of exoplanetary atmospheres, providing new insights into the formation of cool giant planets. 

In this study, we present an extensive parameter space survey of the prominence of H$_2$O in reflection spectra of cool giant planets. We explore the influence of a wide range of effective temperatures, gravities, metallicities, and sedimentation efficiencies, providing a grid of $>$ 50,000 models for the community. In what follows, we espouse the principal components of reflection spectra models in \S\ref{sec:reflection_spectra}. We illustrate how H$_2$O signatures are influenced by planetary parameters in \S\ref{sec:parameters}. We identify how to maximise the detectability of H$_2$O signatures in cool giant planets in \S\ref{sec:max_detectability}. The implications for observations and promising target planets for direct imagining spectroscopy are presented in \S\ref{sec:implications}. Finally, in \S\ref{sec:discussion}, we summarise our results and discuss wider implications. 

\begin{figure*}[ht!]
\epsscale{1.10}
\plotone{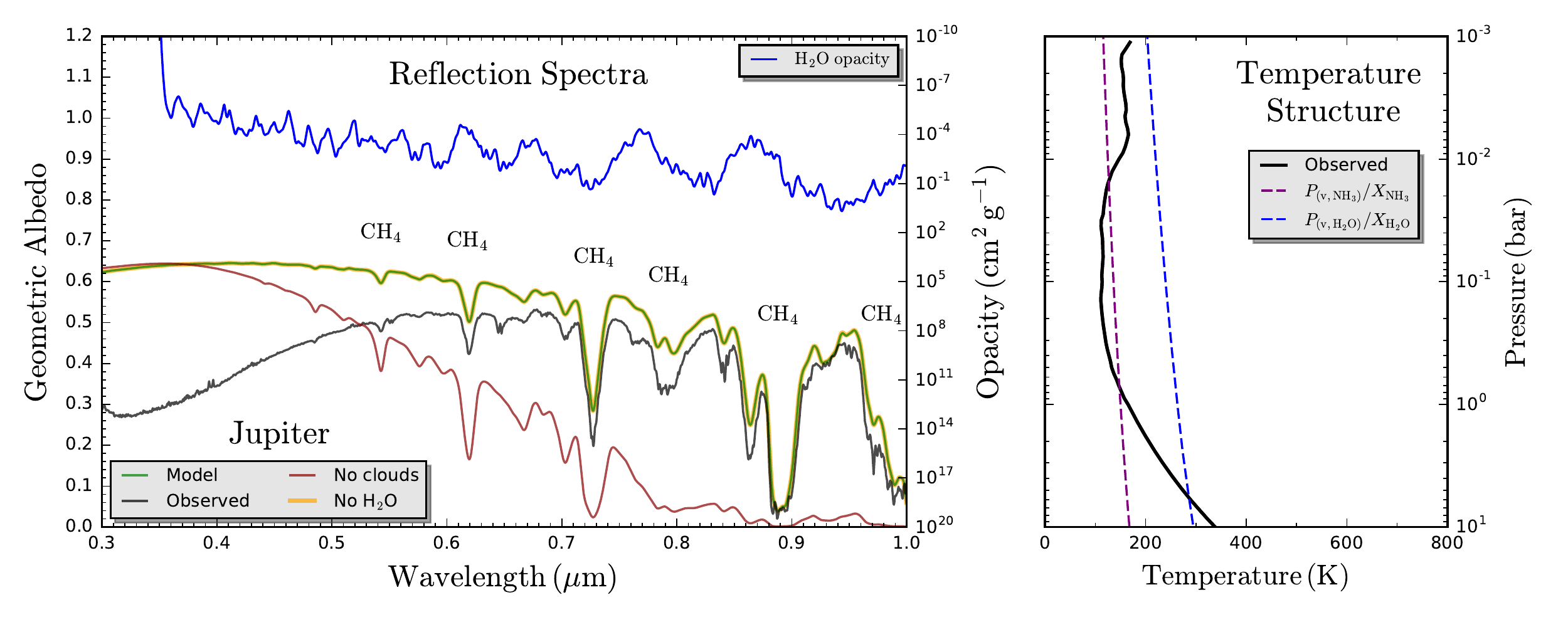}
\label{fig:Jupiter_spectra}
\caption{Jovian geometric albedo spectra and temperature structure. Left: comparison of reflection spectra models including clouds (green), without clouds (red), without H$_2$O vapour (orange), and the observed disk-integrated spectrum of \citet{Karkoschka1994} (black). Prominent CH$_4$ absorption features are annotated. The spectrum with H$_2$O vapour removed is identical to that with H$_2$O present, indicating no H$_2$O signatures are present for Jupiter. The H$_2$O opacity is shown in blue (smoothed for clarity). Right: Voyager-derived temperature structure of Jupiter (black) \citep{Lindal1981}, with the NH$_3$ (purple) and H$_2$O (blue) condensation curves for a $3 \times$solar atmosphere.}
\end{figure*}

\newpage

\section{Reflection Spectroscopy of Exoplanetary Atmospheres} \label{sec:reflection_spectra}

We begin with a brief exploration of reflection spectroscopy, focusing specifically on the physical processes that control their appearance. We qualitatively illustrate these features through consideration of Jupiter's reflection spectrum, before elaborating on the computation of reflection spectra for exoplanetary atmospheres. 

\subsection{Geometric Albedo Spectra}

Consider a fully-illuminated planetary disk. A fraction of the stellar flux will be scattered by the planetary atmosphere, enabling one to define the \emph{geometric albedo spectrum}:

\begin{equation}
A_{g}(\lambda) = \left(\frac{a}{R_{p}}\right)^{2} \frac{F_{p} \, (\alpha=0, \lambda)}{F_{*}(\lambda)}
\label{eq:geometric_albedo}
\end{equation}
where $a$ is the planet-star orbital separation, $R_p$ is the planetary radius, $F_{p} \, (\alpha=0, \lambda)$ is the observed scattered flux from the fully-illuminated planetary disk (the planet at opposition in solar system parlance), and $F_{*} (\lambda)$ is the stellar flux. $\alpha$ is the phase angle, defined as the angle between the stellar ray incident of the planet and the  ray towards the observer. For transiting exoplanets, the condition of full illumination ($\alpha = 0$)  occurs upon star-planet osculation (i.e. secondary eclipse). In practise, observations of reflection spectra will be taken during partial illumination ($\alpha \neq \, 0$), such that

\begin{equation}
\frac{F_{p} \, (\alpha, \lambda)}{F_{*}(\lambda)} = A_{g}(\lambda) \left(\frac{R_{p}}{a}\right)^{2} \Phi \, (\alpha, \lambda)
\label{flux_ratio}
\end{equation}
where the decrease in scattered flux is encoded in the phase function, $\Phi \, (\alpha, \lambda)$ -- normalised to $1$ when $\alpha = 0$.

When discussing reflection spectra in this work, we express our results in terms of geometric albedo spectra, $A_{g}(\lambda)$. We note that the expression given in Equation \ref{eq:geometric_albedo} is entirely equivalent to the historical definition of geometric albedo as the ratio of the observed scattered flux from the planet at full illumination to that of an isotropically scattering disk of the same cross sectional area as the planet and placed at the same distance \citep[][Chapter 3]{Seager_book}. For future work, it would be beneficial to supplant this historical terminology with more intuitive nomenclature better suited to dealing with reflection spectra of exoplanets.

The shape of a geometric albedo spectrum is primarily controlled by the relative contributions of molecular absorption opacity, molecular Rayleigh scattering, and the scattering opacity of condensate clouds. The cloud properties, in turn, are determined by the temperature structure and metallicity of the atmosphere. Radiative transfer of stellar photons impinging on the atmosphere then enables determination of the scattered flux, $F_{p}(\alpha=0, \lambda)$, and hence the geometric albedo spectrum. Before elaborating on the details of such a calculation, we first turn to illustrate how these atmospheric properties influence the geometric albedo spectrum of Jupiter.

\subsection{Geometric Albedo Spectra of Jupiter}

Figure \ref{fig:Jupiter_spectra} shows the observed disk-averaged albedo spectrum of Jupiter \citep{Karkoschka1994}, along with three models demonstrating the influence of clouds and H$_2$O vapour. The models assume a 3$\times$ solar abundance atmosphere, with the Jovian temperature structure derived from Voyager \citep{Lindal1981}. The spectra of both the models and observations can be broken down into a scattering continuum with embedded absorption features. Especially prominent CH$_4$ features are noted at 0.54$\micron$, 0.62$\micron$, 0.73$\micron$, 0.79$\micron$, 0.89$\micron$, and 0.99$\micron$ \citep{Karkoschka1994,Cahoy2010}. The cloud-free model has a continuum provided by Rayleigh scattering, whilst the models with clouds exhibit a significantly brighter continuum at longer wavelengths due to back-scattering from NH$_3$ clouds. This brightening substantially additionally enhances the contrast of absorption features at longer wavelengths -- an example of clouds enhancing the detection of molecular species in reflection spectra.

Comparing our models to observations, we see that the observed spectrum is generally darker at shorter wavelengths. This is due to our models not including the influence of photochemical hazes present in Jupiter's upper atmosphere \citep{Marley1999,Sudarsky2000}. Smaller discrepancies at longer wavelengths are likely due to different cloud properties and deviations between Jupiter's molecular abundances and the uniform 3$\times$ solar abundance ratio we assume. Nevertheless, the general qualitative agreement without parameter tuning to fit the observed spectrum serves as a satisfactory demonstration of the efficacy of our modelling framework.

\subsubsection{Jupiter's Concealed H$_2$O}

To assess the prominence of H$_2$O vapour absorption in Jupiter's reflection spectrum, Figure \ref{fig:Jupiter_spectra} also shows how the model with clouds changes when all H$_2$O vapour is removed. If H$_2$O absorption signatures were present in the spectrum, this would result in a higher albedo in regions of high H$_2$O opacity. Instead, we find this model is everywhere coincident with the model including H$_2$O, demonstrating that no H$_2$O absorption features are imprinted in our model Jupiter spectrum. Regarding observational signatures, \citet{Karkoschka1994} noted a slight decrease in Jupiter's albedo relative to Saturn between 0.92 - 0.95 $\micron$, coincident with the maximum H$_2$O opacity in the optical, suggesting that this could potentially be attributed to $\mathrm{H_{2}O}$ absorption. However, this feature was later ascribed to $\mathrm{NH_{3}}$ \citep{Karkoschka1998}. Thus our models and disk-averaged observations are in concordance that Jupiter displays no prominent H$_2$O signatures.

To understand why this is the case, Jupiter's temperature structure \citep{Lindal1981} is displayed on the right of Figure \ref{fig:Jupiter_spectra} against condensation curves for $\mathrm{NH_{3}}$ \citep[Appendix A]{Ackerman2001} and $\mathrm{H_{2}O}$ \citep{Buck1981}. In a simple conceptual picture, a species existing in vapour form in the deep atmosphere begins to condense once the partial pressure it exerts exceeds the species' saturation vapour pressure

\begin{equation}
P \, (T) \, X_{v, \, i} \geq P_{\mathrm{vap}, \, i} \, (T)
\label{eq:condensation_pressure}
\end{equation}
where $P \, (T)$ is the atmospheric pressure in a given layer, $X_{v, \, i}$ is the vapour mixing ratio of species $i$, and $P_{\mathrm{vap}, \, i} \, (T)$ is the corresponding vapour pressure of species $i$. As the saturation vapour pressure sharply decreases with temperature (Figure \ref{fig:Jupiter_spectra}, higher layers satisfy this condition once the partial pressure equals the saturation vapour pressure. Above this point, the species begins to condense and the vapour mixing ratio, decreases such that equality is maintained in Equation \ref{eq:condensation_pressure}. This condition is equivalent to finding the deepest layer in which the pressure-temperature (P-T) profile intersects the condensation curve, $P_{\mathrm{vap}, \, i} \, (T) / X_{v, \, i}$. Note that the mixing ratio causes condensation curves to shift to higher temperatures as metallicity increases. If no intersection occurs, the species will remain in gaseous form throughout the atmosphere. For Jupiter, Figure \ref{fig:Jupiter_spectra} demonstrates that intersection occurs for both NH$_3$ and H$_2$O, forming clouds at pressures of $\sim$0.8 bar and $\sim$5 bar respectively, matching well with observations \citep{Atreya1999}. H$_2$O clouds thus form sufficiency deep that rainout depletes the observable atmosphere of detectable quantities of $\mathrm{H_{2}O}$ vapour. The resulting albedo spectrum is then dominated by the higher-altitude NH$_3$ clouds and absorption due to CH$_4$ vapour. Jupiter's concealed H$_2$O serves to demonstrate the crucial importance played by the atmospheric temperature structure, which governs both the altitudes where clouds form and the types of cloud present.

\subsubsection{Revealing H$_2$O Absorption Signatures}

From examining Jupiter's temperature structure (Figure \ref{fig:Jupiter_spectra}), one can conceptualise how the broad characteristics of reflection spectra alter for cool giant exoplanets. Consider an exoplanet marginally warmer than Jupiter, such that $T_{\mathrm{eff}} \approx 150$ K. Such a minor perturbation is sufficient to ensure the NH$_3$ condensation curve is no longer crossed, with the H$_2$O condensation curve intersection occurring higher in the atmosphere. The anticipated result is an atmosphere whose geometric albedo at longer wavelengths is dominated by reflection from H$_2$O clouds. Due to the rising H$_2$O cloud base as a function of temperature, it is expected that the fraction of photons reaching the cloud deck, and hence reflected towards the observer, will increase with temperature. These reflected photons will encounter abundant $\mathrm{H_{2}O}$ vapour on their inbound and outbound trajectories, resulting in observable H$_2$O absorption signatures. This `cloud enhancement' must exhibit a limiting temperature, as the $\mathrm{H_{2}O}$ clouds themselves will begin to dissipate for even warmer planets. However, this limit must also be metallicity dependant, as the condensation curve for a given pressure shifts to higher temperature for greater deep mixing ratios (Equation \ref{eq:condensation_pressure}), allowing H$_2$O clouds to persist to higher temperatures. Additional factors, such as the gravitational field strength (altering the gradient of the P-T profile in the deep atmosphere) and the sedimentation efficiency (controlling cloud thickness and formation altitude) will also influence the prominence of H$_2$O signatures. Our objective in this study is to quantify the relative contributions of each of these factors. With a conceptual picture of the principal physical processes established, we now elaborate on the calculation of reflection spectra.

\subsection{Modelling Exoplanet Reflection Spectra} \label{subsec:methods}

\begin{figure}[t!]
\epsscale{1.20}
\plotone{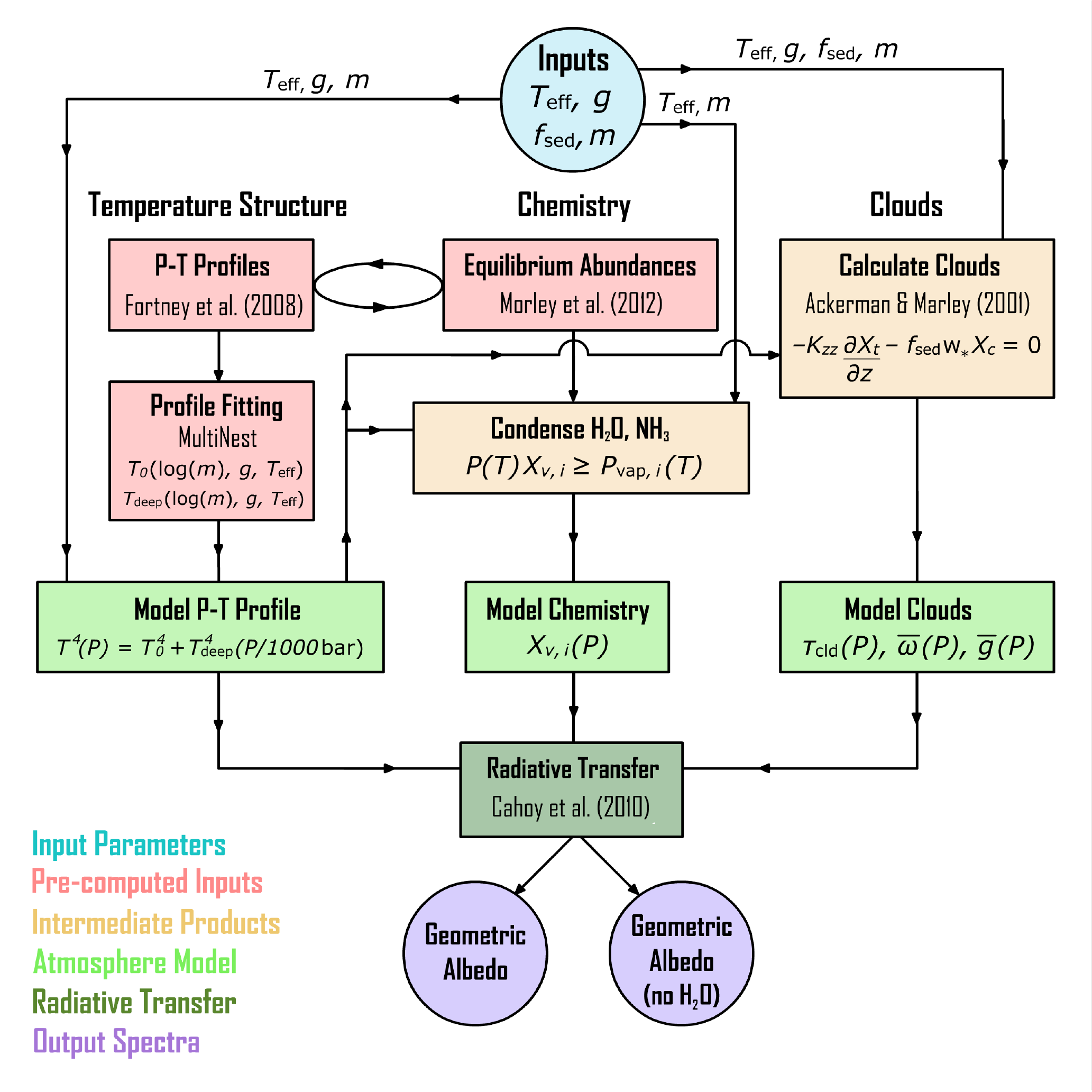}
\label{fig:flowchart}
\caption{Schematic diagram of our reflection spectra modelling procedure. Arrows indicate operation order, with the loop representing self-consistent iterative coupling. Radiative-convective equilibrium pressure-temperature (P-T) profiles \citep{Fortney2008} and thermochemical equilibrium abundances \citep{Morley2012} were pre-computed self-consistently for a grid of cloud-free planets in gravity-distance-metallicity space. A simple parametric function was fit to the equilibrium P-T profiles, yielding $T(P)$ as a function of $g$, $m$, and $T_{\mathrm{eff}}$. Where $T(P)$ cross a condensation curve, the equilibrium abundances of condensates are updated due to rainout. For a given $T(P)$, $m$, $f_{\mathrm{sed}}$, the \citet{Ackerman2001} cloud model yields the cloud optical depth, single scattering albedo, and asymmetry parameter in each layer. The temperature structure, chemical abundances, and cloud properties are used to solve the equation of radiative transfer \citet{Cahoy2010}, yielding two geometric albedo spectra for each value of $g$, $T_{\mathrm{eff}}$, $f_{\mathrm{sed}}$, and $m$ -- one with and one without H$_2$O vapour opacity.}
\end{figure}

Computing a reflection spectrum requires specification of three atmospheric properties: (i) the pressure-temperature profile; (ii) chemical abundances; and (iii) cloud characteristics and optical properties. Given these inputs, the equation of radiative transfer is solved according to the geometry at a specified orbital phase (here, $\alpha = 0$). Figure \ref{fig:flowchart} illustrates the steps involved in our modelling process, each of which we examine in turn.

\subsubsection{Temperature Structure} \label{subsubsec:PT_profile}

Given a specified metallicity, $m$, C/O ratio, gravitational field strength, $g$, internal temperature, $T_{\mathrm{int}}$, and incident flux, $F_{*}$, one may iteratively calculate the pressure-temperature structure (P-T profile) of an atmosphere. Such `self-consistent' calculations usually assume radiative-convective equilibrium and chemical equilibrium. In principle, one temperature profile can be computed for each desired point in parameter space \citep[as in][]{Cahoy2010}, though this approach is computationally expensive and can sometimes fail to reach a converged solution, particularly if clouds are included as part of the iteration step \citep{Morley2014}. In this study, where our goal is to compute reflection spectra across a wide swath of parameter space, such an approach is not computationally feasible.

Our approach is to develop an \emph{a priori} determined P-T profile that is able to reproduce the behaviour of self-consistent profiles. We fit a simple two-parameter function of the form 

\begin{equation}
T^{4} (P) = T_{0}^{4} + T_{\mathrm{deep}}^{4} (P/1000 \mathrm{bar})
\label{eq:PT_parametrisation}
\end{equation}
where $T_{0}$ and $T_{\mathrm{deep}}$ encode the temperatures of an upper-atmosphere isotherm and the temperature at the 1000 bar level, respectively, to a range of self-consistent P-T profiles computed using the formalism of \citet{Fortney2008}. The profiles we consider span $m$ from 1$\times$ to 100$\times$ solar, $g$ from 10 ms$^{-2}$ to 100 ms$^{-2}$, $a$ from 0.5 AU to 5.0 AU, and all have $T_{\mathrm{int}}$ = 150 K and C/O = 0.5. We note that these profiles assume a cloud-free atmosphere. From integrating the thermal emission, we derive the effective temperature, $T_{\mathrm{eff}}$, of each atmosphere. From the totality of these fits, we expand $T_{0}$ and $T_{\mathrm{deep}}$ as functions of metallicity, gravity, and effective temperature, such that we arrive at analytic expressions for $T_{0} (\mathrm{log}(m), \, g, \, T_{\mathrm{eff}})$ and $T_{\mathrm{deep}} (\mathrm{log}(m), \, g, \, T_{\mathrm{eff}})$. The fitting procedure, parameter functional forms, and fitting coefficients are detailed in Appendix \ref{Appendix}.

\begin{figure*}[ht!]
\epsscale{1.12}
\plotone{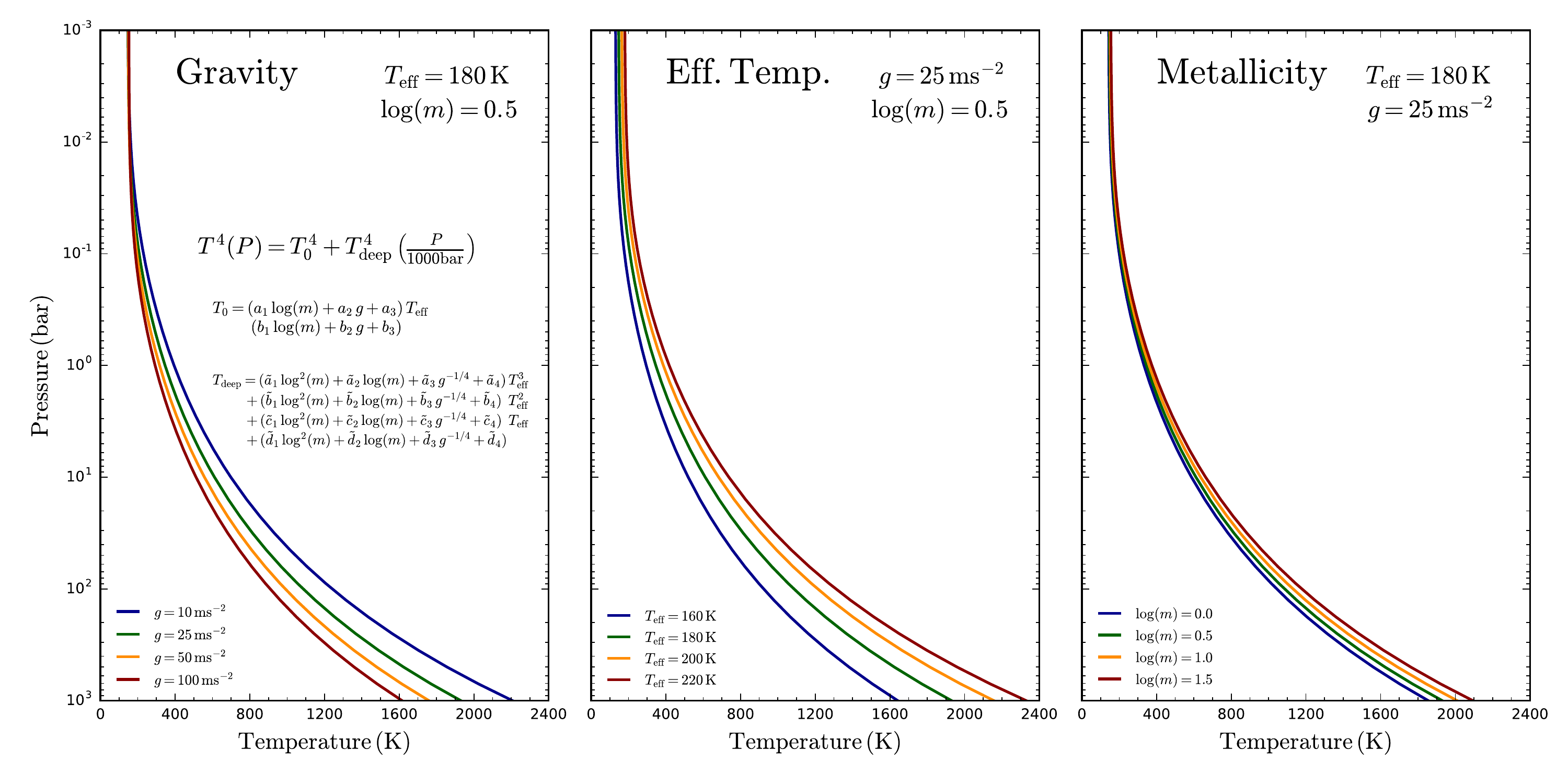}
\label{fig:PT_param_plot}
\caption{A new analytic P-T profile for cool giant planets. From left to right, the panels show how the temperature structure changes with $g$, $T_{\mathrm{eff}}$, and $m$ for a reference model with $g$ = 25 ms$^{-2}$, $T_{\mathrm{eff}}$ = 180 K, and m = 3$\times$ solar. All models assume $T_{\mathrm{int}}$ = 150 K, and do not explicitly account for clouds. The analytic profile was derived through fitting a grid of radiative-convective equilibrium P-T profiles \citep{Fortney2008} to a simple parametric function, as detailed in Appendix \ref{Appendix}.}
\end{figure*}

The variation of our analytic P-T profile with $g$, $T_{\mathrm{eff}}$, and $\mathrm{log}(m)$ is shown in Figure \ref{fig:PT_param_plot}. The broad trends are as follows: (i) lower gravity planets exhibit sharper adiabats in the deep atmosphere (due to convective onset occurring at lower pressures); (ii) higher effective temperature planets have higher isothermal temperatures and sharper adiabats; (iii) higher metallicity planets are slightly warmer (due to the additional molecular opacity), but with similar slope adiabats. The simple functional form of our profile allows the computational bottleneck of radiative-convective equilibrium calculations to be circumvented, and hence enables efficient exploration of an extensive region of $g$-$T_{\mathrm{eff}}$-$m$ parameter space. 

\subsubsection{Atmospheric Chemistry} \label{subsubsec:chemistry}

At the temperatures relevant to cool giant planets, atmospheres are dominated by H$_2$ and He, with the principal oxygen, carbon, and nitrogen reservoirs being H$_2$O, CH$_4$, and NH$_3$ \citep{Burrows1997, Madhusudhan2016}. Our models consider a wide range of chemical species: H$_2$, He, H+, H-, H$_2$O, CH$_4$, NH$_3$, CO, CO$_2$, PH$_3$, H$_2$S, N$_2$, Na, K, TiO, VO, and Fe. The opacities were computed via the k-coefficient method \citep{Freedman2008}, updated for ammonia \citep{Yurchenko2011} and $\mathrm{H_2}$ collisionally-induced opacity \citep{Richard2012} as described in \citet{Saumon2012}. For each of the \citet{Fortney2008} P-T profiles used in the previous section, the abundances of all species were calculated self-consistently assuming thermochemical equilibrium, according to the prescription in \citet{Morley2012}. For our purposes, we take the chemistry at each point in parameter space to be representative of the nearest point in the $m$-$T_{\mathrm{eff}}$ plane from the grid of self-consistent models.

As the chemical equilibrium abundances are derived for cloud-free atmospheres, we modify the mixing ratios of $\mathrm{NH_{3}}$ and $\mathrm{H_{2}O}$ to account for rainout. As a first order correction, we update the vapour mixing ratios following \citet{Lewis1969}, starting at the base of the atmosphere and assuming all material in excess of the vapour pressure condenses, such that

\begin{equation}
X_{v, \, i} \, (z) = \text{min} [X_{v, \, i} \, (z - \Delta z), \, P_{\mathrm{vap}, \, i} \, (z)/P \, (z)]
\label{eq:condensation_Lewis}
\end{equation}
where $z$ specifies the atmospheric layer. The vapour pressures for H$_2$O and NH$_3$ are from \citet{Buck1981} and \citet{Ackerman2001}, respectively. This correction is crucial to ensure a degree of consistency between the chemical abundances employed in the geometric albedo calculation and the generated cloud profile (section \ref{subsubsec:cloud_model}). Despite this correction, we note that the usage of a stand-alone cloud model is not strictly self-consistent with the chemical abundances, despite using the same P-T profile for both calculations. However, our cloud model features a more sophisticated treatment of condensation than Equation \ref{eq:condensation_Lewis}, as we shall now see.

\subsubsection{Cloud Model} \label{subsubsec:cloud_model}

We model H$_2$O and NH$_3$ clouds via the prescription of \citet{Ackerman2001}. This model balances the upwards flux of condensate and vapour due to turbulent mixing with the downwards flux of condensate due to sedimentation. For each condensing species, this implies

\begin{equation}
-K_{zz} \frac{\partial X_{t}}{\partial z} - f_{\mathrm{sed}}\, w_{*}\, X_{c} = 0 
\label{eq:Ackerman_Marley}
\end{equation}
where $K_{zz}$ is the vertical eddy diffusion coefficient, $X_{t} = X_{v} + X_{c}$ is the total combined species mixing ratio of vapour and condensate, $w_{*}$ is the convective velocity scale, and $f_{\mathrm{sed}}$ is a tunable parameter called the \emph{sedimentation efficiency}. Physically, greater values of $f_{\mathrm{sed}}$ lead to larger condensate particle sizes with more efficient rainout, and hence thinner cloud profiles. $K_{zz}$ and $w_{*}$ are both determined via mixing length theory, assuming the diffusion coefficient corresponding to free convective heat transport \citep{Gierasch1985} is equal to $K_{zz}$. In convectively stable regions, a minimum value of $K_{zz} = 10^{5} \,\mathrm{cm}^{2}\, \mathrm{s}^{-1}$ is prescribed to account for sources of residual turbulence in radiative regions. With $f_{\mathrm{sed}}$ and the atmospheric metallicity, $m$ (controlling the deep mixing ratios of H$_2$O and NH$_3$), specified, Equation \ref{eq:Ackerman_Marley} is solved for the condensate mixing ratio in each layer, assuming that all excess vapour condenses (Equation \ref{eq:condensation_pressure}).

Once the condensate mixing ratios are determined, the next step is the specification of a condensate particle size distribution. For this purpose, we assume each condensed species follows a lognormal distribution:

\begin{equation}
\frac{dn}{dr} = \frac{N}{r \sqrt{2 \pi}\, \mathrm{ln} \sigma_{g}} e^{-\frac{\mathrm{ln^{2}}(r/r_{g})}{2 \mathrm{ln^{2}} \sigma_{g}}}
\label{eq:lognormal}
\end{equation}
where $n(r)$ is the number density of condensate particles with radii $<r$, $N$ is the total number density for a given condensate, $r_{g}$ is the geometric mean radius, and $\sigma_{g}$ is the geometric standard deviation (here set to be 2.0 throughout). Full specification of this distribution requires determination of $r_{g}$, accomplished by interpreting $f_{\mathrm{sed}}\, w_{*}$ as the mass-weighted sedimentation velocity:
\begin{equation}
f_{\mathrm{sed}}\, w_* = \frac{\displaystyle \int_{0}^{\infty} v_{f}(r)\, r^{3} \frac{dn}{dr} dr}{\displaystyle \int_{0}^{\infty} r^{3} \frac{dn}{dr} dr}
\label{eq:sedimentation_velocity}
\end{equation}
where $v_{f}(r)$ is the particle `fall' speed at a given radius - calculated assuming viscous flow around spherical particles and corrected for kinetic effects \citep[see][Appendix B]{Ackerman2001}. Substituting Equation \ref{eq:lognormal} into Equation \ref{eq:sedimentation_velocity} and evaluating the integrals then yields $r_{g}$. $N$ can similarly be obtained by integrating Equation \ref{eq:lognormal} which, together with $r_{g}$, fully specifies the distribution in terms of the only remaining free parameter, $f_{\mathrm{sed}}$.

With the size distribution determined, the cloud scattering properties are calculated using Mie theory. We assume spherical, homogeneous, particles. The complex refractive indices of H$_2$O ice \citep{Warren1984} and NH$_3$ ice \citep{Martonchik1984} are utilised to evaluate the Mie efficiencies at 2000 uniformly-spaced wavelengths from 0.3 - 1.0 $\micron$. The efficiencies are then integrated over the particle size distribution in Equation \ref{eq:lognormal} to produce the single scattering albedo, $\tilde{\omega}$, asymmetry parameter, $\tilde{g}$, and cloud optical depth, $\tau_{\mathrm{cld}}$, in each atmospheric layer as a function of wavelength.

\subsubsection{Radiative Transfer} \label{subsubsec:radiative_transfer}

Having specified the temperature structure, chemical abundances, and cloud properties as a function of altitude, the equation of radiative transfer is solved to compute the scattered flux from the planetary atmosphere. We evaluate the geometric albedo using the approach developed by \citet{Toon1977,Toon1989,McKay1989,Marley1999,Marley1999a} and extended to arbitrary phase angles by \citet{Cahoy2010}. In this approach, the planetary hemisphere is divided into a number of plane-parallel facets. Each of these facets represents an atmospheric column in which 1D radiative transfer, including multiple scattering, is evaluated. We calculate radiative transfer as described in \citet{McKay1989} and \citet{Marley1999}. Other applications and further details can be found in \citet{McKay1989,McKay1989a,Marley1996,Burrows1997,Marley1999a,Marley2002,Saumon2008,Fortney2008,Cahoy2010}. We assume the same P-T profile, chemistry, and cloud properties are present in each column -- effectively representing hemispheric average properties. However, since each facet experiences different angles for both the incident stellar flux and the angle to the observer, the scattered flux will vary with latitude and longitude. The observed geometric albedo is then produced via Chebyshev-Gauss integration over the hemisphere's two dimensional coordinate system to average over viewing geometry \citep[see][for greater detail]{Cahoy2010}. We typically evaluate geometric albedo spectra at 2000 wavelengths, uniformly-spaced from 0.3 to 1.0 $\micron$.

This framework gives the flexibility to explore an extensive parameter space. Our P-T profiles are directly specified as functions of $g$, $T_{\mathrm{eff}}$, and $\mathrm{log}(m)$; chemistry as a function of $m$ and $T_{\mathrm{eff}}$; and cloud properties as a function of $m$ and $f_{\mathrm{sed}}$. The chemistry and cloud properties are also indirectly influenced by the other parameters through the shape of the P-T profile (see Figure \ref{fig:flowchart}). We can thus model reflection spectra of exoplanetary atmospheres spanning the 4-dimensions of $T_{\mathrm{eff}}$, $g$, $m$, and $f_{\mathrm{sed}}$. For each point in parameter space, we generate two geometric albedo spectra: one following the prescription as described above, and a second where we artificially remove all H$_2$O vapour opacity during the radiative transfer calculation. It is the differences between these two models that enables us to assess the prominence of H$_2$O absorption in the atmosphere.

\section{Factors Influencing H$_2$O Prominence in Cool Giant Planets} \label{sec:parameters}

We begin our investigation into H$_2$O signatures in cool giant planets by examining the spectral regions in which they occur and how their prominence is affected by planetary parameters. We consider, in turn, the impact of effective temperature, gravity, sedimentation efficiency, and metallicity. Specifically, we take a reference giant planet model with $T_{\mathrm{eff}}$ = 180 K, $g$ = 10 ms$^{-2}$, $f_{\mathrm{sed}}$ = 3, and $m$ = 3$\times$ solar, examining how H$_2$O signatures change as each parameter is perturbed. We choose this reference model to be $\sim$ 60 K warmer than Jupiter, such that NH$_3$ clouds do not form and hence weak H$_2$O signatures are already visible for such a planet.

\begin{figure*}[ht!]
\centering
\includegraphics[width=0.995\textwidth]{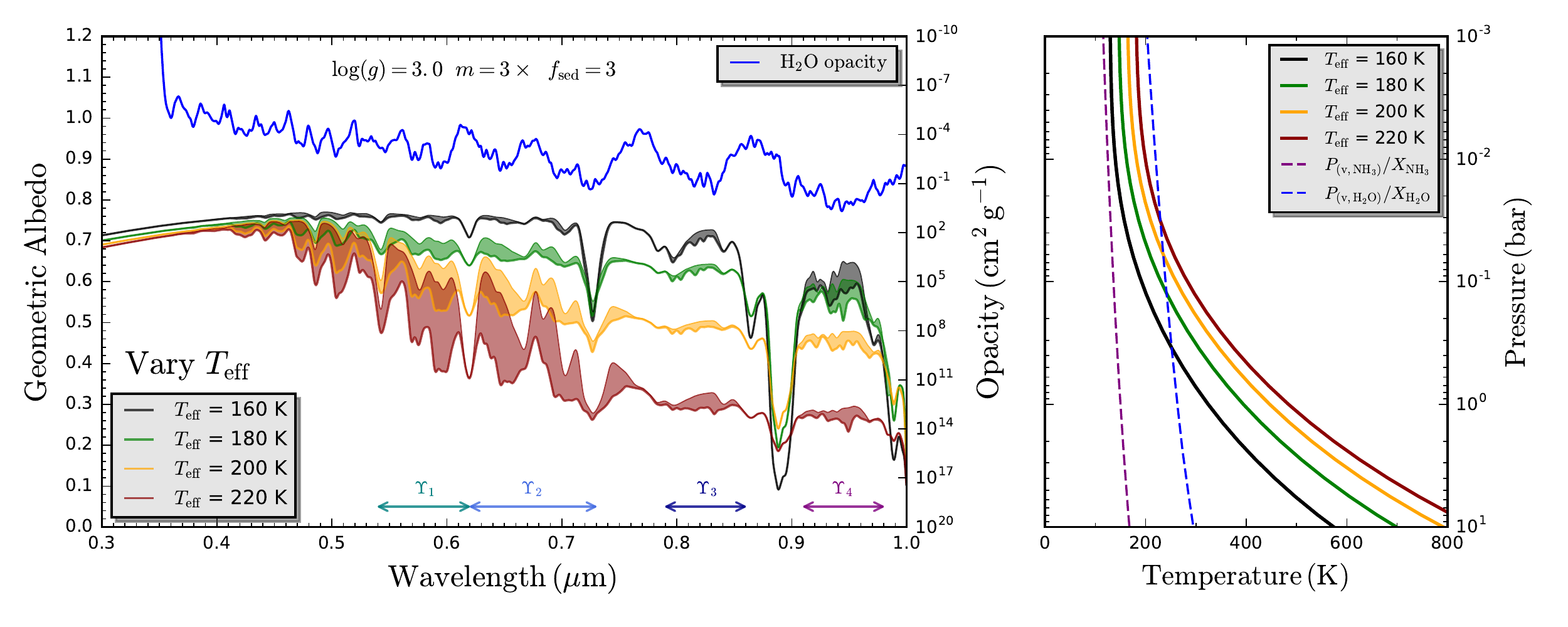}
\label{fig:vary_Teff}
\caption{The influence of effective temperature on H$_2$O absorption signatures in reflection spectra of cool giant planets. Left: geometric albedo spectra models of four planets, with $T_{\mathrm{eff}}$ = 160 K (black), 180 K (green), 200 K (orange), and 220 K (red). All planets have $g$ = 10 ms$^{-2}$ ($\mathrm{log}(g) \,  (\mathrm{cgs})$ = 3.0, $m$ = 3$\times$ solar, and $f_{\mathrm{sed}}$ = 3. The green model is a common reference in Figures \ref{fig:vary_Teff}, \ref{fig:vary_g}, \ref{fig:vary_m}, and \ref{fig:vary_fsed}. Two models are plotted for each planet: one with H$_2$O opacity enabled (lower curve), and one without H$_2$O opacity (upper curve), with the region between the models shaded. The shaded region thus indicates H$_2$O absorption signatures. The H$_2$O opacity is shown in blue (smoothed for clarity). Four spectral regions with prominent H$_2$O signatures are indicated by $\Upsilon_{1,2,3,4}$ (see Section \ref{subsec:quantify_indices}). Right: P-T profiles corresponding to the same four planet models, with the NH$_3$ (purple) and H$_2$O (blue) condensation curves for a $3 \times$solar atmosphere.}
\end{figure*}

\begin{figure*}[ht!]
\centering
\includegraphics[width=0.995\textwidth]{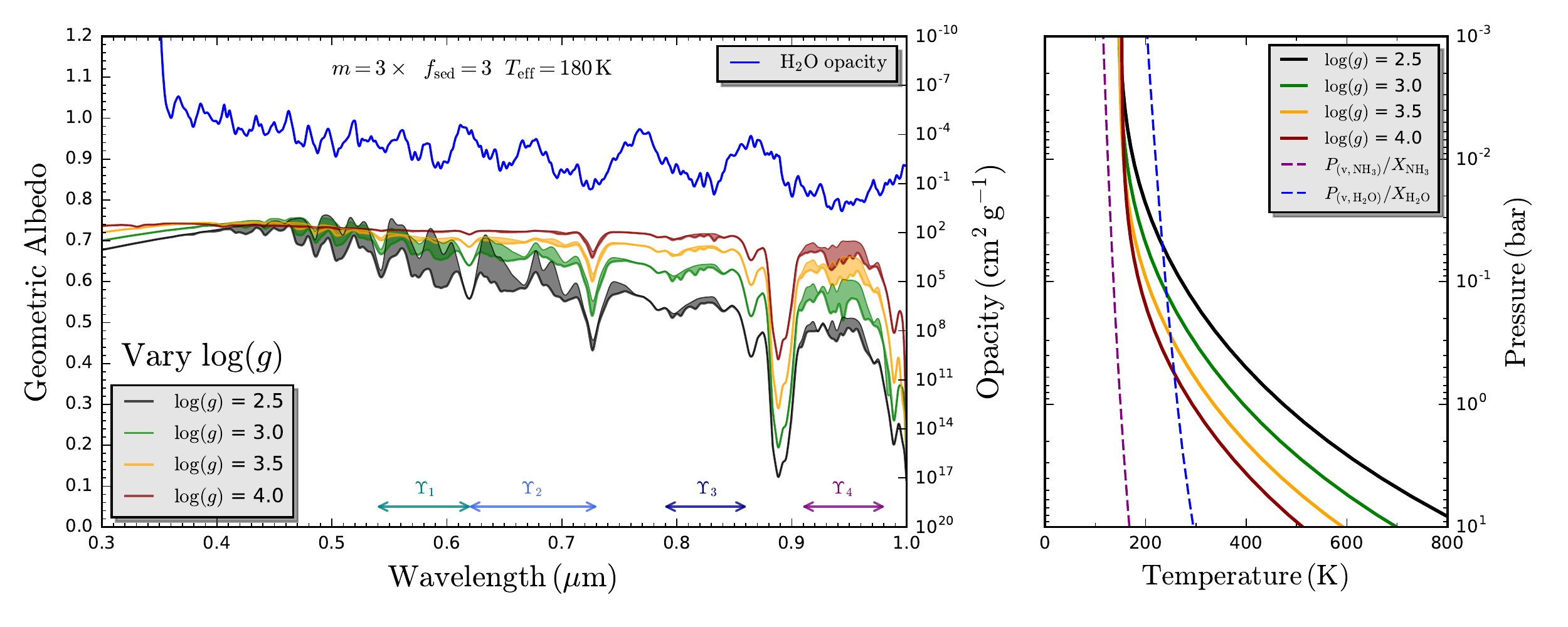}
\label{fig:vary_g}
\caption{The influence of gravity on H$_2$O absorption signatures in reflection spectra of cool giant planets. Left: geometric albedo spectra models of four planets, with $\mathrm{log} (g) \, (\mathrm{cgs})$ = 2.5 (black), 3.0 (green), 3.5 (orange), and 4.0 (red) -- corresponding to $g$ =  3.16, 10.0, 31.6, and 100.0 ms$^{-2}$, respectively. All planets have $m$ = 3$\times$ solar, $f_{\mathrm{sed}}$ = 3, and $T_{\mathrm{eff}}$ = 180 K. The green model is a common reference in Figures \ref{fig:vary_Teff}, \ref{fig:vary_g}, \ref{fig:vary_m}, and \ref{fig:vary_fsed}. Two models are plotted for each planet: one with H$_2$O opacity enabled (lower curve), and one without H$_2$O opacity (upper curve), with the region between the models shaded. The shaded region thus indicates H$_2$O absorption signatures. The H$_2$O opacity is shown in blue (smoothed for clarity). Four spectral regions with prominent H$_2$O signatures are indicated by $\Upsilon_{1,2,3,4}$ (see Section \ref{subsec:quantify_indices}). Right: P-T profiles corresponding to the same four planet models, with the NH$_3$ (purple) and H$_2$O (blue) condensation curves for a $3 \times$solar atmosphere.}
\end{figure*}

\subsection{Effective Temperature} \label{subsec:params_Teff}

Figure \ref{fig:vary_Teff} demonstrates how changes in $T_{\mathrm{eff}}$ affect geometric albedo spectra of cool giant planets. In general, cooler planets are brighter, especially at longer wavelengths, whilst warmer planets are darker at most wavelengths. H$_2$O absorption features are seen even for the coolest model shown ($T_{\mathrm{eff}}$ = 160 K). Prominent H$_2$O absorption is seen from $\sim$ 0.91 - 0.98 $\micron$ and $\sim$ 0.79 - 0.86 $\micron$, coinciding with the first and second maxima of the H$_2$O optical opacity, and across a forest of features from $\sim$ 0.4 - 0.73 $\micron$. These regions are indicated in Figure \ref{fig:vary_Teff} by $\Upsilon_{1,2,3,4}$, where we have split the `H$_2$O forest' into two regions, taking the CH$_4$ features at 0.54, 0.62, and 0.73 $\micron$ as natural dividers. The prominence of H$_2$O absorption in these regions will be quantitatively explored in terms of spectral indices in Section \ref{sec:max_detectability}. H$_2$O absorption features at $<$ 0.54 $\micron$ may suffer from darkening due to photochemical hazes (see Figure \ref{fig:Jupiter_spectra}), leading to an overstatement of their prominence in our model spectra, so for this reason we do not assign these features a separate spectral index.

The strength of H$_2$O absorption features can vary dramatically with temperature. For $\lambda \lesssim$ 0.8 $\micron$ H$_2$O signatures are strongest for high $T_{\mathrm{eff}}$, whilst the 0.94 $\micron$ feature is strongest for lower $T_{\mathrm{eff}}$. The feature around 0.83 $\micron$ represents an intermediate case that is relatively insensitive to $T_{\mathrm{eff}}$. These trends are caused by $\tau_{\mathrm{cld}}$ decreasing for planets with greater $T_{\mathrm{eff}}$, due to the P-T profile intersecting the H$_2$O condensation curve at higher altitudes where the lower pressure results in less mass of H$_2$O being available to condense (Equation \ref{eq:condensation_pressure}). Specifically, the peak values of the cloud optical depth, $\tau_{\mathrm{cld, max}}$, are 74, 22, 8, and 3 for $T_{\mathrm{eff}}$ = 160 K, 180 K, 200 K, and 220 K, respectively. For low temperature planets the albedo is thus dominated by the highly reflective ($\bar{\omega} >0.999$) H$_2$O clouds, which provide a backscattering surface for photons at long wavelengths. The H$_2$O feature around 0.94 $\micron$ is the most prominent for such cool planets due to its greater opacity. At higher temperatures, the optically thin clouds are less effective at reflecting photons and hence more are lost to the deep atmosphere, resulting in both a darker planet and a weaker 0.94 $\micron$ feature. However, the H$_2$O features for $\lambda \lesssim$ 0.8 $\micron$ increase in prominence at higher $T_{\mathrm{eff}}$ due to Rayleigh scattering providing a cloud-independent avenue for short wavelength photons to probe the deep atmosphere, accrue absorption features, and emerge unscathed.

We note in passing that many of these trends can be understood qualitatively by examining the toy cloud model discussed in \citet{Marley2000}. Cloud opacity depends on the available column mass of material that can condense, which itself depends on gravity, temperature, and cloud base pressure.

\subsection{Gravity} \label{subsec:params_g}

Figure \ref{fig:vary_g} demonstrates how changes in $g$ affect geometric albedo spectra. Gravitational field strength contributes to the spectra in two principal ways: (i) changing the P-T profile, thereby affecting cloud formation; (ii) changing the density scale height of the atmosphere. (i) is by far the dominant of these effects, as changes in the cloud optical depth and pressure level controls both the overall reflectivity of the atmosphere and the prominence of H$_2$O features. As higher gravity planets have steeper P-T profiles, they intersect the H$_2$O vapour pressure curve lower in the atmosphere, hence forming deep, optically thick, cloud decks ($\tau_{\mathrm{cld, max}} \approx 50$ for the $\mathrm{log} (g) = 4.0$ (cgs) in Figure \ref{fig:vary_g}). This results in bright albedo spectra with the 0.94 $\micron$ H$_2$O feature most prominent. Planets with low $g$ exhibit `gravity enhanced' H$_2$O features below 0.8 $\micron$ due to Rayleigh scattering dominating when $\tau_{\mathrm{cld}}$ is low -- similar to the trend seen for high $T_{\mathrm{eff}}$ planets in section \ref{subsec:params_Teff}. (ii) generally results in stronger H$_2$O features for lower $g$, as the larger density scale height raises the column abundance of H$_2$O encountered by photons before backscattering at a given pressure level. The compound effect of (i) and (ii) is thus a substantial gravity enhancement in H$_2$O features $<$ 0.8 $\micron$ for planets with $g \lesssim$ 10 ms$^{-2}$.

\subsection{Sedimentation Efficiency} \label{subsec:params_fsed}

\begin{figure*}[ht!]
\includegraphics[width=0.995\textwidth]{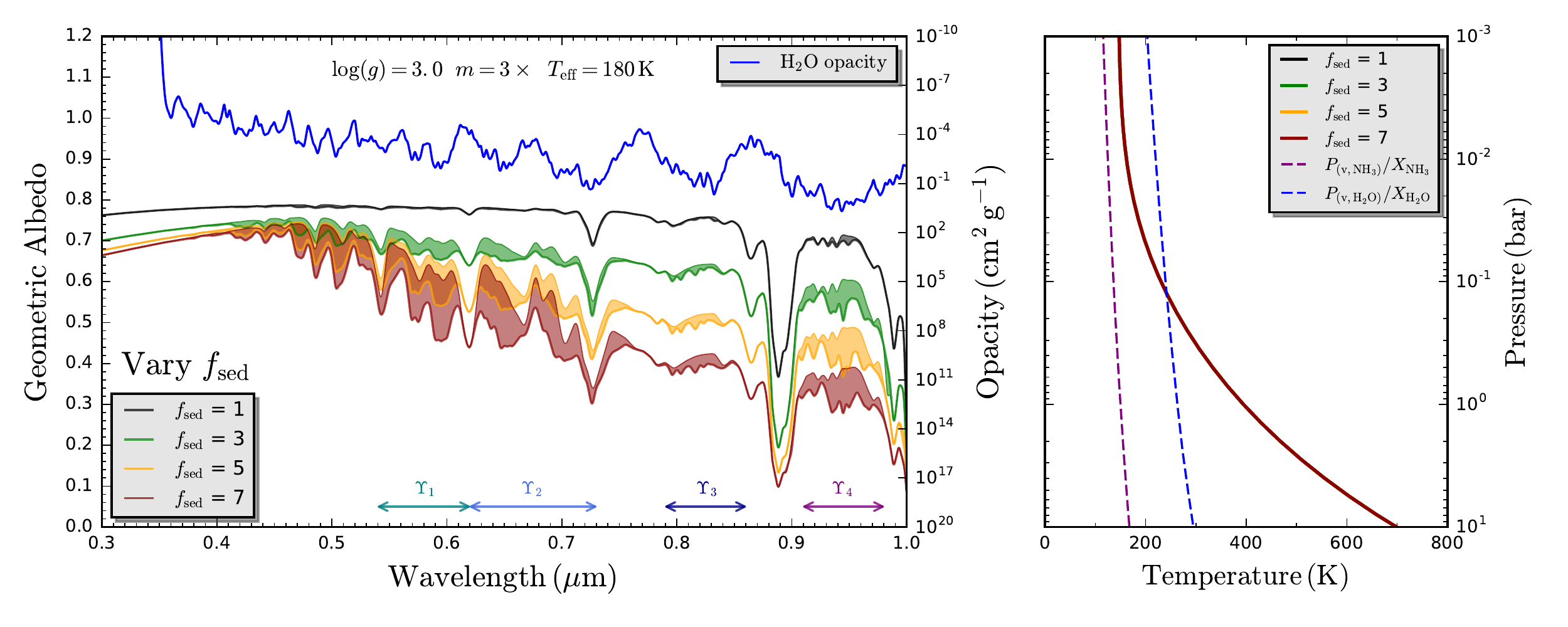}
\label{fig:vary_fsed}
\caption{The influence of sedimentation efficiency on H$_2$O absorption signatures in reflection spectra of cool giant planets. Left: geometric albedo spectra models of four planets, with $f_{\mathrm{sed}}$ = 1 (black), 3 (green), 5 (orange), and 7 (red). All planets have $g$ = 10 ms$^{-2}$ ($\mathrm{log}(g) \,  (\mathrm{cgs})$ = 3.0), $m$ = 3$\times$ solar, and $T_{\mathrm{eff}}$ = 180 K. The green model is a common reference in Figures \ref{fig:vary_Teff}, \ref{fig:vary_g}, \ref{fig:vary_m}, and \ref{fig:vary_fsed}. Two models are plotted for each planet: one with H$_2$O opacity enabled (lower curve), and one without H$_2$O opacity (upper curve), with the region between the models shaded. The shaded region thus indicates H$_2$O absorption signatures. The H$_2$O opacity is shown in blue (smoothed for clarity). Four spectral regions with prominent H$_2$O signatures are indicated by $\Upsilon_{1,2,3,4}$ (see Section \ref{subsec:quantify_indices}). Right: P-T profiles corresponding to the same four planet models, with the NH$_3$ (purple) and H$_2$O (blue) condensation curves for a $3 \times$solar atmosphere. Note that the four P-T profiles are coincident, as our temperature profile is invariant to cloud properties (Section \ref{subsubsec:PT_profile}).}
\end{figure*}

Figure \ref{fig:vary_fsed} demonstrates how changes in $f_{\mathrm{sed}}$ affect geometric albedo spectra. Generally, higher sedimentation efficiencies lead to darker planets with deeper H$_2$O absorption features. Unlike for $T_{\mathrm{eff}}$ and $g$, this is not caused by changes in the location of the cloud base. Clouds begin to form at a common value of $\sim$ 250 mbar for the depicted models, as our P-T profile parametrisation is invariant to $f_{\mathrm{sed}}$ (section \ref{subsubsec:PT_profile}). Rather, the relevant factor here is the cloud thickness above the base, which controls the overall optical depth of the cloud layer. Greater values of $f_{\mathrm{sed}}$ result in larger mean particle radii, leading to more efficient rainout and hence thinner clouds (section \ref{subsubsec:cloud_model}). Specifically, for the four models shown in Figure \ref{fig:vary_fsed}, the cloud vertical extents with $\tau_{\mathrm{cld}} > 1$ in each layer are $\sim$ 20-250 mbar, 50-250 mbar, 60-250 mbar, and 90-250 mbar; for $f_{\mathrm{sed}}$ = 1, 3, 5, and 7, respectively. $\tau_{\mathrm{cld, max}}$ in each case is 84, 22, 12, and 7, occurring around 130 mbar. We see then that the thickest clouds in vertical extent are also the most efficient at scattering, explaining the high albedo for models with low $f_{\mathrm{sed}}$. Models with higher $f_{\mathrm{sed}}$ then display stronger H$_2$O features due to the greater path length traversed before encountering the upper edge of the cloud deck. H$_2$O absorption at 0.94 $\micron$ thus becomes more prominent for higher $f_{\mathrm{sed}}$, but at the cost of a lower continuum albedo due to less photons reaching the clouds when the upper edge is at deeper pressures. However, features below 0.8 $\micron$ have the benefit of an underlying Rayleigh scattering continuum, leading to H$_2$O features which stand out against a bright continuum for high $f_{\mathrm{sed}}$.

\begin{figure*}[ht!]
\includegraphics[width=0.995\textwidth]{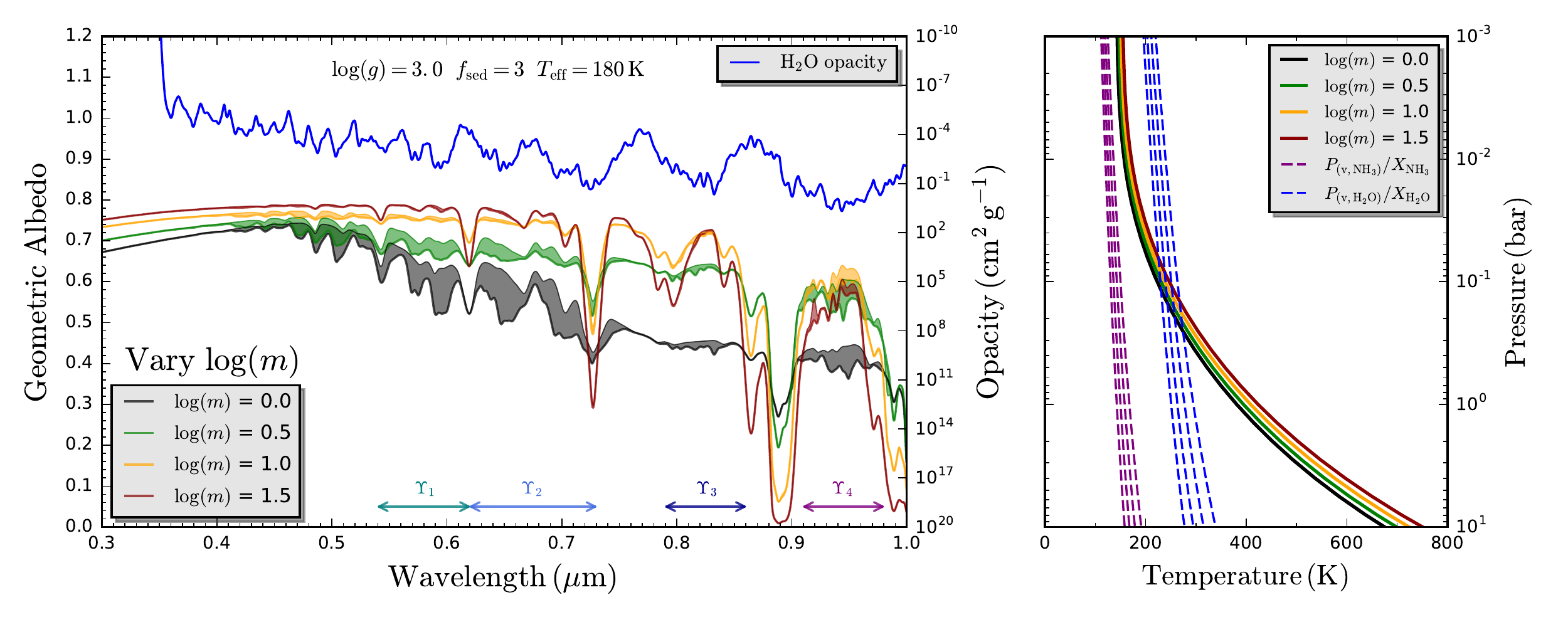}
\label{fig:vary_m}
\caption{The influence of metallicity on H$_2$O absorption signatures in reflection spectra of cool giant planets. Left: geometric albedo spectra models of four planets, with $\mathrm{log} (m)$ = 0.0 (black), 0.5 (green), 1.0 (orange), and 1.5 (red) $\times$ solar -- corresponding to $m$ =  1.00, 3.16, 10.0, and 31.6$\times$ solar, respectively. All planets have $g$ = 10 ms$^{-2}$ ($\mathrm{log}(g) \,  (\mathrm{cgs})$ = 3.0), $f_{\mathrm{sed}}$ = 3, and $T_{\mathrm{eff}}$ = 180 K. The green model is a common reference in Figures \ref{fig:vary_Teff}, \ref{fig:vary_g}, \ref{fig:vary_m}, and \ref{fig:vary_fsed}. Two models are plotted for each planet: one with H$_2$O opacity enabled (lower curve), and one without H$_2$O opacity (upper curve), with the region between the models shaded. The shaded region thus indicates H$_2$O absorption signatures. The H$_2$O opacity is shown in blue (smoothed for clarity). Four spectral regions with prominent H$_2$O signatures are indicated by $\Upsilon_{1,2,3,4}$ (see Section \ref{subsec:quantify_indices}). Right: P-T profiles corresponding to the same four planet models, with the NH$_3$ (purple) and H$_2$O (blue) condensation curves for each metallicity value, with metallicity increasing from left to right.}
\end{figure*}

\subsection{Metallicity} \label{subsec:params_m}

Figure \ref{fig:vary_m} demonstrates how changes in $m$ affect geometric albedo spectra. Three trends emerge as metallicity increases: (i) CH$_4$ absorption features strengthen; (ii) the continuum albedo brightens; (iii) H$_2$O absorption features weaken. The strengthening of CH$_4$ features can be seen most strikingly around 0.73 $\micron$ and 0.89 $\micron$, which is simply a consequence of the increased CH$_4$ abundance for models with higher $m$. At first glance, it is somewhat counter-intuitive that H$_2$O features do not strengthen in the same manner, despite the increased H$_2$O abundance. This difference can be understood in terms of H$_2$O condensation, as CH$_4$ exists in vapour form throughout the atmosphere (condensing only for $T_{\mathrm{eff}} \lesssim 100$ K). First, note that the H$_2$O condensation curves and P-T profiles intersect at roughly the same pressure (Figure \ref{fig:vary_m}, right), so the cloud bases occur at similar altitudes. However, higher $m$ implies higher deep abundances, increasing the H$_2$O partial pressure, and thereby causing an expanded inventory of condensed H$_2$O. The result is the formation of increasingly optically thick clouds ($\tau_{\mathrm{cld, max}}$ = 6, 22, 79, 204, for $\mathrm{log} (m)$ = 0.0, 0.5, 1.0, and 1.5), which form a bright backscattering surface that dominates over all other opacity for high $m$.

\section{Maximising Detectability of H$_2$O in Cool Giant Planets} \label{sec:max_detectability}

Having demonstrated how reflection spectra and H$_2$O features qualitatively change with planetary properties, in this section we proceed to identify regions of parameter space with maximally prominent H$_2$O signatures. We begin by introducing a quantitative measure of the observability of H$_2$O signatures, before utilising this metric to extend the results of the previous section to thousands of model atmospheres. 

\subsection{Quantifying H$_2$O Observability} \label{subsec:quantify_indices}

Signatures of H$_2$O manifest via irregular perturbations to the continuum, prohibiting standard absorption feature metrics such as equivalent width \citep{Chamberlain1987}. Therefore, we develop a quantitative prescription for evaluating the observability of H$_2$O based on two factors: (i) the strength of H$_2$O absorption features; (ii) the continuum albedo inside the absorbing window. A balance must be established between these factors, as even a strong H$_2$O feature will not be observable if it results in a planet sufficiently dark to render albedo measurements infeasible. With this compromise in mind, we propose spectral indices of the form:
\begin{equation}
\Upsilon_{i} = 4 \, \, \overline{\Delta A_{g}} \times \overline{A_{g}}
\label{eq:spec_index_equation}
\end{equation}
where $\overline{\Delta A_{g}} = \left(\overline{A_{g,}}_{\mathrm{\, no \, H_{2}O}} - \overline{A_{g}} \right)$ is the mean difference between the geometric albedos of two models without and with H$_2$O vapour opacity enabled, and $\overline{A_{g}}$ is the mean continuum geometric albedo (with H$_2$O opacity). The factor of 4 is an arbitrary constant set such that $\Upsilon_{i, \, \mathrm{max}} = \left(\overline{A_{g,}}_{\mathrm{\, no \, H_{2}O}} \right)^{2}$, occurring when $\overline{A_{g}} = (1/2) \, \overline{A_{g,}}_{\mathrm{\, no \, H_{2}O}}$. The index $i$ indicates the spectral range over which mean quantities are evaluated, e.g:
\begin{equation}
\overline{\Delta A_{g}} = \frac{1}{\lambda_{2} - \lambda_{1}}  \int_{\lambda_{1}}^{\lambda_{2}} \left[ A_{g, \, \mathrm{\, no \, H_{2}O}} \, (\lambda) - A_{g} \, (\lambda) \right] \, d\lambda
\label{eq:mean_albedo_difference}
\end{equation}
where the pairs of values $(\lambda_{1}, \, \lambda_{2})$ for each spectral index are as follows: $\Upsilon_{1} \in (0.54 \, \micron, \, 0.62\, \micron)$; $\Upsilon_{2} \in (0.62 \, \micron, \, 0.73\, \micron)$; $\Upsilon_{3} \in (0.79 \, \micron, \, 0.86\, \micron)$; $\Upsilon_{4} \in (0.91 \, \micron, \, 0.98\, \micron)$.

\begin{figure}[ht!]
\includegraphics[width=0.95\columnwidth]{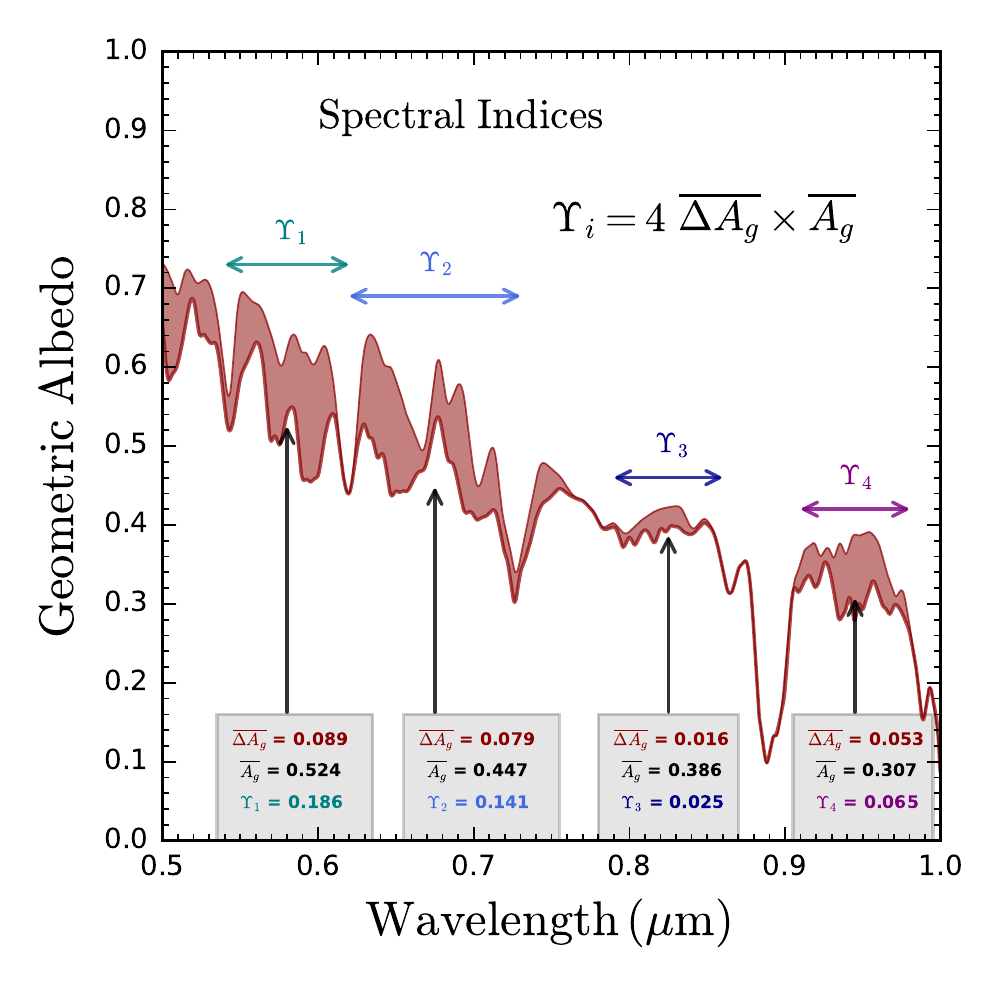}
\label{fig:spec_indices}
\vspace{-0.5cm}
\caption{A quantitative prescription for evaluating H$_2$O prominence in reflection spectra. The bold red curve shows the geometric albedo spectrum of a model with $g$ = 10 ms$^{-2}$, $m$ = 3 $\times$ solar, $f_{\mathrm{sed}}$ = 7, and $T_{\mathrm{eff}}$ = 180 K. The light red curve shows the geometric albedo of an identical model with H$_2$O vapour opacity disabled, such that the enclosed red shaded region is caused by H$_2$O absorption. Four spectral regions with prominent H$_2$O features are indicated. The black arrows show the mean geometric albedo, $\overline{A_{g}}$, within each spectral region. Spectral indices $\Upsilon_{i}$ are calculated for each region using Equations \ref{eq:spec_index_equation} and \ref{eq:mean_albedo_difference}, resulting in the values in the boxes.}
\end{figure}

We demonstrate the calculation of the spectral indices $\Upsilon_{i}$ for a model reflection spectrum in Figure \ref{fig:spec_indices}. This figure also provides a convenient geometrical picture of the terms in Equation \ref{eq:spec_index_equation}. $\overline{\Delta A_{g}}$ is simply the coloured area within a given spectral region divided by the width of the region. $\overline{A_{g}}$, shown by the black arrows, provides a measure of the continuum albedo which would be observed within a given spectral region. As the enclosed area increases (due to greater H$_2$O absorption), the black arrows necessarily decrease, resulting in their product exhibiting a maximum value when $\overline{\Delta A_{g}} = \overline{ A_{g}} = (1/2) \, \overline{A_{g,}}_{\mathrm{\, no \, H_{2}O}}$ (i.e. the shaded region has a width-averaged height equal to the black arrow). We thus see that $\Upsilon_{i}$ favours planets with deep H$_2$O features that also posses a reasonable continuum albedo, enshrining the desired compromise wherein observable H$_2$O features arise.

\subsection{Exploring H$_2$O Prominence}

We now proceed to explore the prominence of H$_2$O features across a wide range of atmospheric models. We saw in Section \ref{sec:parameters} that H$_2$O signatures are influenced by changes in $T_{\mathrm{eff}}$, $g$, $f_{\mathrm{sed}}$, and $m$, and, crucially, that a parameter change that enhances H$_2$O features at short wavelengths can simultaneously weaken them at long wavelengths. With spectral indices on hand, we shall generalise these conclusions by systematically computing the prominence of H$_2$O features in multiple spectral regions for an ensemble of models.

We consider the 4-dimensional parameter space spanned by $T_{\mathrm{eff}}$, $g$, $f_{\mathrm{sed}}$, and $m$. Our models range from $T_{\mathrm{eff}} = 150 \rightarrow 400$ K, $\mathrm{log} (g) = 2.0 \rightarrow 4.0$ (cgs), $f_{\mathrm{sed}} = 1 \rightarrow 10$, and $\mathrm{log} (m) = 0.0 \rightarrow 2.0 \times$ solar. We discretise this parameter space into intervals of $\Delta T_{\mathrm{eff}}$ = 10 K, $\Delta \mathrm{log} (g)$ = 0.1 dex, $\Delta f_{\mathrm{sed}}$ = 1, and $\Delta \mathrm{log} (m)$ = 0.5 dex, generating reflection spectra both with and without H$_2$O opacity enabled via the methodology outlined in Section \ref{subsec:methods}. The resulting 54,600 geometric albedo spectra are provided for the community\footnote{\href{https://doi.org/10.5281/zenodo.1210305}{Reflection Spectra Repository}}.

The reasoning behind the chosen parameter space is as follows. Our coolest models are $\sim 30$ K warmer than Jupiter, set by the requirement that $T_{\mathrm{eff}} > T_{\mathrm{int}}$ (our P-T profiles are based on models with $T_{\mathrm{int}} = 150$ K -- see Appendix \ref{Appendix}). The warmest models ($T_{\mathrm{eff}} \sim 400$ K) correspond to $a \sim 0.5$ AU around a solar analogue, sufficient to cover exoplanets explorable by the inner working angles anticipated of upcoming direct imaging missions. The range of $g$ from 1 ms$^{-2} \rightarrow$ 100 ms$^{-2}$ was chosen to include a wide range of possible surface gravities for radial-velocity planets with known values of $M \mathrm{sin}(i)$. Similarly, $m$ was chosen to range from $1 \rightarrow 100 \times$ solar to encompass a wide range of possible atmospheric metallicities. Typical values of $f_{sed}$ considered to date have varied from 1-3 for planets \citep{Ackerman2001}, up to 6 for brown dwarfs \citep{Stephens2009}, and as high as 10 for some cool giant planets \citep{Cahoy2010}. Values as small as $f_{\mathrm{sed}} = 0.01$ have also been considered to explain flat transmission spectra \citep{Kreidberg2014,Morley2015}. We chose $f_{sed}$ = 1 as a minimum boundary, as smaller values produce thick highly-extended clouds that lead to uniformly bright albedo spectra with few absorption features \citep{Morley2015}. The maximum value of $f_{sed}$ = 10 was chosen to encompass physically motivated plausible values. 

\begin{figure*}[ht!]
\centering
\includegraphics[width=0.95\textwidth]{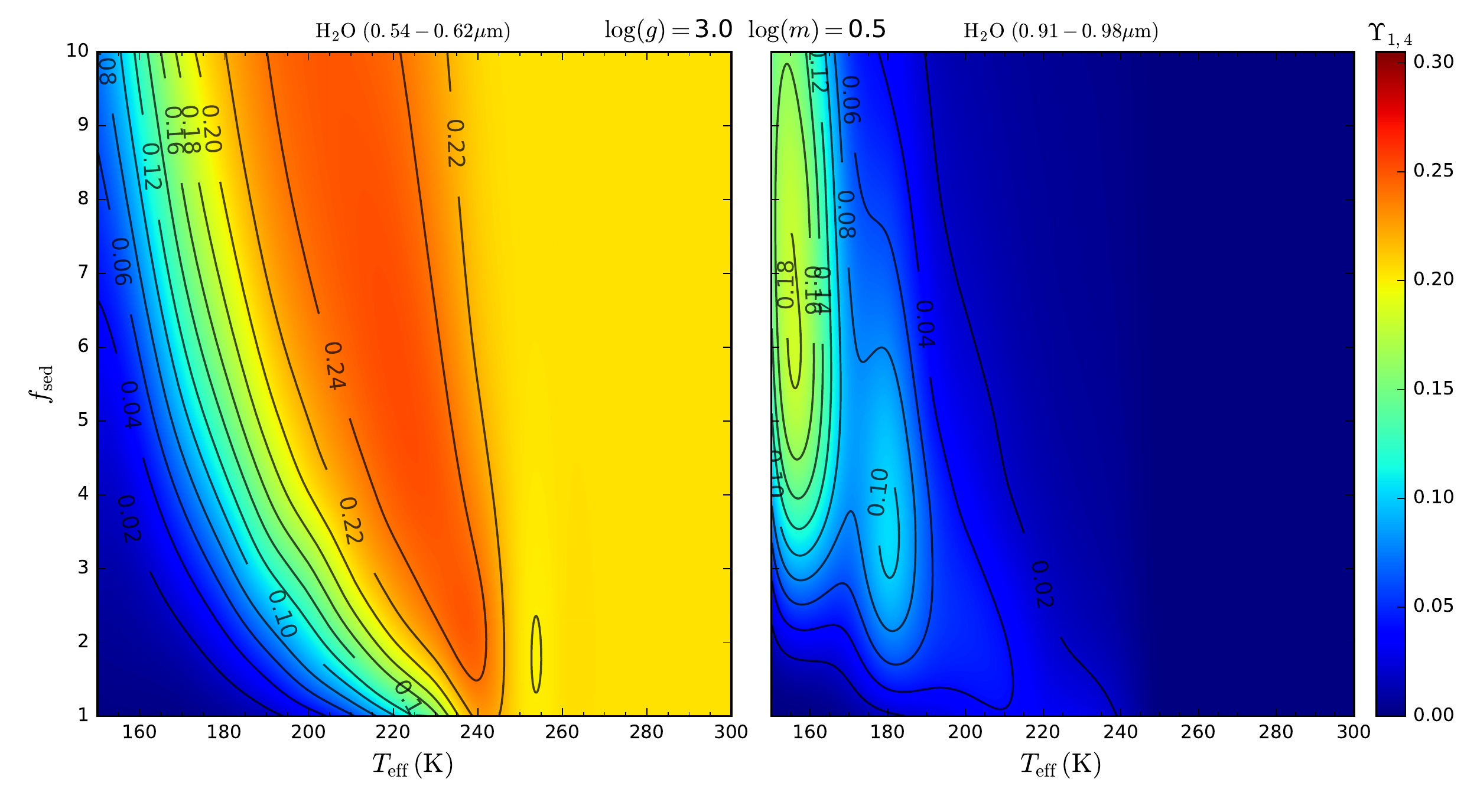}
\label{fig:contour_f-T}
\caption{H$_2$O prominence in the $f_{\mathrm{sed}}$ - $T_{\mathrm{eff}}$ plane. Left: H$_2$O absorption on the Rayleigh slope ($\sim$ 0.6 $\micron$). Right: H$_2$O absorption at the optical opacity maximum ($\sim$ 0.94 $\micron$). The colouring indicates the values of the spectral indices $\Upsilon_{1}$ (left) and $\Upsilon_{4}$ (right) -- see section \ref{subsec:quantify_indices}. The variation with $g$ and $m$ are shown in the animated figure, available in the HTML article.}
\end{figure*}

The prominence of H$_2$O signatures across this 4-dimensional parameter space is shown in Figure 9. We demonstrate how the observability of H$_2$O varies across a series of $f_{\mathrm{sed}}$ - $T_{\mathrm{eff}}$ planes, quantified by the spectral indices $\Upsilon_{1}$ (left) and $\Upsilon_{4}$ (right). $\Upsilon_{1}$ represents H$_2$O signatures along the Rayleigh scattering slope ($\sim 0.6 \, \micron$), whilst $\Upsilon_{4}$ represents H$_2$O signatures around the optical opacity maximum ($\sim 0.94 \, \micron$). We do not show $\Upsilon_{2}$ or $\Upsilon_{3}$, as the former displays the same trends as $\Upsilon_{1}$, whilst the latter tends to be less prominent than the other spectral indices (e.g., Figure \ref{fig:spec_indices}). Each $f_{\mathrm{sed}}$ - $T_{\mathrm{eff}}$ plane contains 160 values of $\Upsilon_{1}$ or $\Upsilon_{4}$, interpolated using a rectangular bivariate spline of cubic order to produce contour maps. The blue regions contain negligible H$_2$O absorption, whilst the red correspond to highly prominent H$_2$O features. The variation with $g$ and $m$ are shown in the animated version of Figure \ref{fig:contour_f-T} (available online in the HTML representation of this study), whilst the static version is for a representative planet with $g$ = 10 ms$^{-2}$ and $m = 3 \times$ solar.

\begin{figure*}[ht!]
\centering
\includegraphics[width=0.95\textwidth]{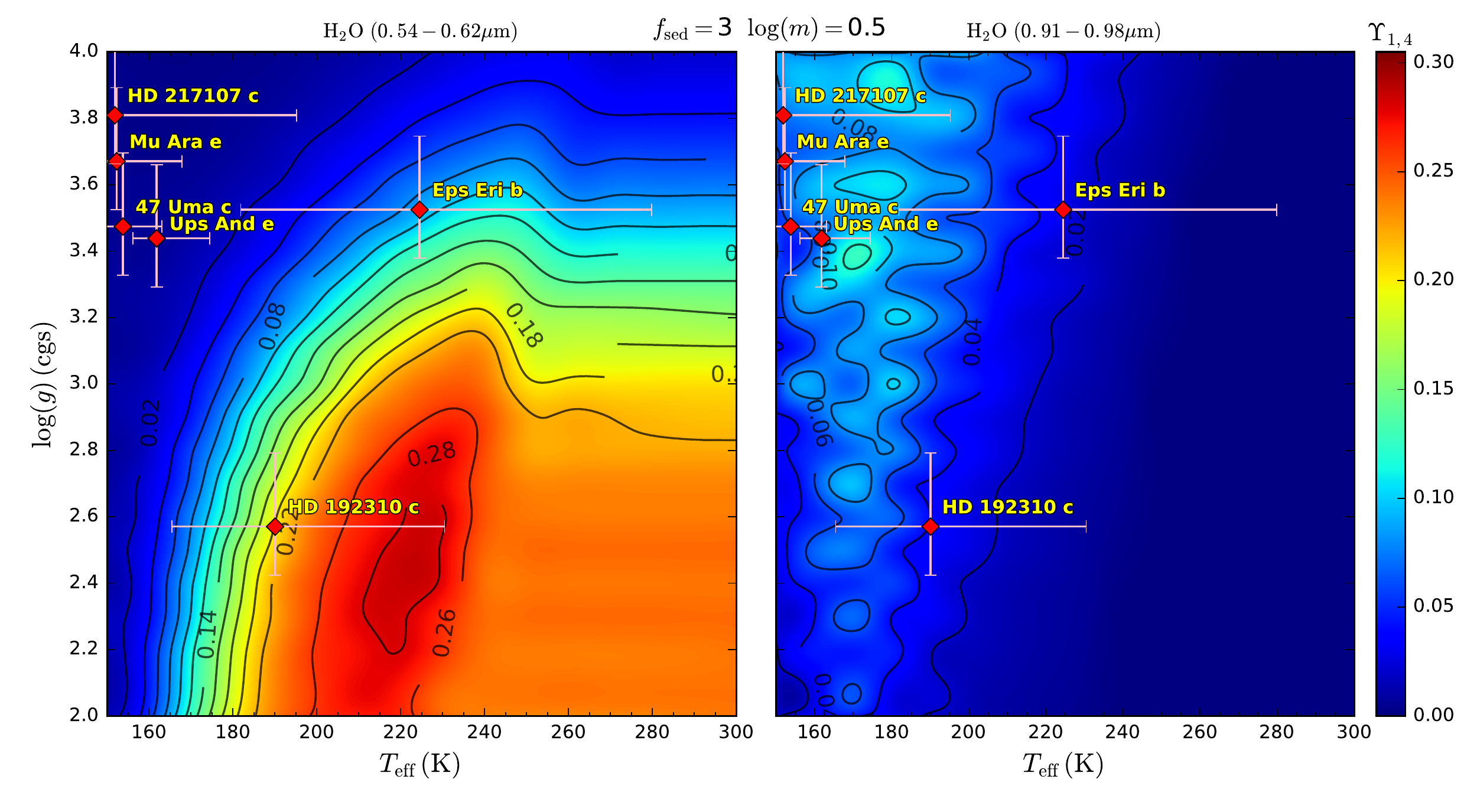}
\label{fig:contour_g-T}
\caption{H$_2$O prominence in the $\mathrm{log} (g)$ - $T_{\mathrm{eff}}$  plane. Left: H$_2$O absorption on the Rayleigh slope ($\sim$ 0.6 $\micron$). Right: H$_2$O absorption at the optical opacity maximum ($\sim$ 0.94 $\micron$). The colouring indicates the values of the spectral indices $\Upsilon_{1}$ (left) and $\Upsilon_{4}$ (right) -- see section \ref{subsec:quantify_indices}. Approximate locations of known radial velocity exoplanets appropriate for direct imaging are shown. The value of $g$ for each planet is obtained by assuming $M_p \mathrm{sin}(i) \sim M_p$ and inferring $R_p$ using mass-radii relations \citep{Fortney2007}. An uncertainty in $g$ of 40$\%$ is ascribed based on these assumptions. The $T_{\mathrm{eff}}$ values and uncertainties are derived from giant planet cooling models \citep{Fortney2008a}, assuming a 100$\%$ error in the system age. The variation with $f_{\mathrm{sed}}$ and $m$ are shown in the animated figure, available in the HTML article.}
\end{figure*}

Two distinct avenues exist to maximise H$_2$O prominence: (i) planets with deep, highly reflective, clouds observed at long wavelengths; (ii) planets with weak, optically thin, clouds at short wavelengths. The difference primarily stems from the wavelength dependence of Rayleigh scattering. Long wavelength H$_2$O features have strong H$_2$O opacity but little Rayleigh scattering, thus requiring clouds to provide a continuum. Short wavelength features can rely on a Rayleigh continuum, but their lower H$_2$O opacity necessitates longer path lengths to accrue H$_2$O signatures; thus favouring optically thin clouds. These two regions can be seen in Figure \ref{fig:contour_f-T}, with an `island' of prominent long wavelength H$_2$O absorption from avenue (i) on the right, and a vast plateau of prominent short wavelength H$_2$O absorption from avenue (ii) on the left. The distinction between these two regions is primarily due to $T_{\mathrm{eff}}$. Low $T_{\mathrm{eff}}$ ($\sim 150$ K) results in the deep optically thick clouds favoured by $\Upsilon_{4}$, whilst $T_{\mathrm{eff}} \gtrsim 180$ K results in weaker clouds favoured by $\Upsilon_{1}$. A maximum in $\Upsilon_{1}$ occurs just prior to the dissipation of H$_2$O clouds ($T_{\mathrm{eff}} \sim 230$ K), caused by cloud backscattering enhancing the continuum albedo above that of Rayleigh scattering (favoured by the second term in Equation \ref{eq:spec_index_equation}). This `cloud enhanced' region is a general feature of $\Upsilon_{1}$ and $\Upsilon_{2}$. Higher values of $f_{\mathrm{sed}}$, resulting in thinner and deeper clouds, tends to enhance both $\Upsilon_{1}$ and $\Upsilon_{4}$ due to the increased path length traversed before scattering occurs. As the observability of $\Upsilon_{4}$ is critically dependant on clouds, values of $f_{\mathrm{sed}} \gtrsim 3$ are typically required to ensure prominent long wavelength features. For larger values of $f_{\mathrm{sed}}$ than considered here (>10), $\Upsilon_{4}$ rapidly falls as efficient particle sedimentation produces optically-thin clouds, leaving planets dark at long wavelengths (e.g. Figure \ref{fig:vary_fsed}).

Metallicity influences the prominence of both short and long wavelength H$_2$O signatures across the entire $f_{\mathrm{sed}}$ - $T_{\mathrm{eff}}$ plane. Generally speaking, $\Upsilon_{1}$ and $\Upsilon_{4}$ rise from solar metallicity until $\mathrm{log} (m) \sim 0.5 \times$ solar, then steadily fall as $m$ rises further. The initial rise in prominence is caused by the increased atmospheric H$_2$O vapour abundance (and brighter continuum due to enhanced H$_2$O cloud scattering), with the decrease caused by high cloud optical depths dominating spectra (e.g. Figure \ref{fig:vary_m}). The regions of maximal H$_2$O prominence shift to higher $T_{\mathrm{eff}}$ as $m$ increases, due to shifting H$_2$O condensation curves. The $\Upsilon_{1}$ `cloud-enhanced' region shifts and contracts from $T_{\mathrm{eff}} \sim 210-240$ K $\rightarrow \, \sim 240-250$ K for $m = 1 \rightarrow 100 \times$ solar and $f_{\mathrm{sed}} \sim 3$, whilst the $\Upsilon_{4}$ `H$_2$O island' shifts and expands from $T_{\mathrm{eff}} \sim 150-160$ K $\rightarrow \, \sim 200-250$ K (Figure \ref{fig:contour_f-T}, animation).

Gravitational field strength plays a crucial role in determining the relative prominences of short wavelength and long wavelength H$_2$O signatures. $\Upsilon_{1}$ is essentially 0 for $g \sim$ 100 ms$^{-2}$, but rapidly rises as $g$ lowers. $\Upsilon_{4}$ instead tends to rise for higher $g$, with the maximal values of $\Upsilon_{1}$ and $\Upsilon_{4}$ in the $f_{\mathrm{sed}}$ - $T_{\mathrm{eff}}$ plane reaching parity at $g \approx 20$ ms$^{-2}$. This differing behaviour is due to low gravity planets forming optically thin cloud decks, as seen in Section \ref{subsec:params_g} (the $\rm CH_4$ clouds of Uranus being a typical example). This dichotomy presents an intriguing possibility: planets with $g \lesssim $ 20 ms$^{-2}$ should display H$_2$O features that are substantially more detectable at wavelengths $< 0.8 \, \micron$ than around the 0.94 $\micron$ H$_2$O feature. This conclusion holds true over a vast parameter space, relatively independent of $f_{\mathrm{sed}}$ and $m$, and for $T_{\mathrm{eff}} \gtrsim 180$ K when $f_{\mathrm{sed}} \lesssim 6 $ (Figure \ref{fig:contour_f-T}, animation). On the other hand, planets with $g \gtrsim $ 20 ms$^{-2}$ may offer better prospects of detecting H$_2$O around 0.94 $\micron$ -- if the values of $T_{\mathrm{eff}}$, $f_{\mathrm{sed}}$, and $m$ are right to enable deep, optically thick, clouds to form. With these trends elucidated, we now turn to examine how these insights can be utilised to inform future direct imaging observations.

\section{Implications for Observations: H$_2$O via Direct Imaging of Exoplanets} \label{sec:implications}

Our exploration of parameter space in the previous section revealed that prominent H$_2$O signatures can exist in reflection spectra of cool giant planets. We shall now link our results to observations, identifying known radial velocity exoplanets amenable to detecting H$_2$O via direct imaging spectroscopy. We conclude by presenting self-consistent reflection spectra of the most promising target exoplanet.

\subsection{Promising Exoplanets for Direct Imaging Spectroscopy} \label{subsec:targets}

To place our results in the context of known exoplanets, we recast our H$_2$O prominence survey in the $\mathrm{log} (g)$ -- $T_{\mathrm{eff}}$ plane, shown in Figure \ref{fig:contour_g-T}. Approximate locations of known radial velocity exoplanets suitable for direct imagining \citep{Lupu2016} are overplotted. As the surface gravities have not been directly measured for these worlds, we infer them by assuming the measured $M_p \, \mathrm{sin}(i) \sim M_p$, and deriving $R_p$ using theoretical mass-radii relations \citep{Fortney2007}. Given the crudeness of these assumptions, we ascribe a 40\% error to each value of $g$. The values of $T_{\mathrm{eff}}$ for each planet are derived from giant planet cooling models \citep{Fortney2008a}, with $T_{\mathrm{eff}}$ error bars accounting for a 100$\%$ error in the system age. As $f_{\mathrm{sed}}$ and $m$ are not \emph{a priori} determined, `Jupiter-like' values of $f_{\mathrm{sed}}$ = 3 \citep{Ackerman2001} and $m = 3 \times$ solar \citep{Wong2004} are presented in the static version of the figure, with the online animated version showing the variability with $f_{\mathrm{sed}}$ and $\mathrm{log} (m)$.

High gravity ($g \gtrsim$ 20 ms$^{-2}$) and low temperature ($T_{\mathrm{eff}} \sim 160$ K) planets are best observed around 0.94 $\micron$. As shown on the right of Figure \ref{fig:contour_g-T}, these conditions best apply to HD 217107 c, Mu Arae e, 47 Ursae Majoris c, and Upsilon Andromedae e. 55 Cancri d (not shown) is also an excellent target, with $T_{\mathrm{eff}} \sim 150$ K and $g \sim$ 100 ms$^{-2}$ falling slightly above the maximum value of $\mathrm{log} (g)$ our models explored. These planets are predicted to possess especially prominent H$_2$O absorption for $f_{\mathrm{sed}} \gtrsim 6$ and $m \lesssim 3 \times$ solar (Figure \ref{fig:contour_g-T}, animation). For higher values of $m$, the `H$_2$O island' these planets reside on shifts to higher temperatures, which could result in Epsilon Eridani b (and possibly HD 192310 c) having detectable H$_2$O features for $m \gtrsim 10 \times$ solar.

Low gravity ($g \lesssim$ 20 ms$^{-2}$) planets are best observed at wavelengths $\lesssim$ 0.8 $\micron$. The left panel of Figure \ref{fig:contour_g-T} demonstrates that the only target planet for which this currently applies is HD 192310 c. Such low gravity planets with $T_{\mathrm{eff}} \gtrsim 180$ K are predicted to have Rayleigh slopes dominated by H$_2$O absorption, far exceeding the prominence of features around 0.94 $\micron$. Higher values of $f_{\mathrm{sed}}$ shift the `H$_2$O plateau' to lower temperatures, improving the prospects of observing H$_2$O on cool, low gravity worlds -- with $f_{\mathrm{sed}} \gtrsim 7$ placing HD 192310 c inside the `cloud enhanced' region of maximum H$_2$O prominence (Figure \ref{fig:contour_g-T}, animation). This effect could also enable Epsilon Eridani b to have observable H$_2$O signatures at short wavelengths for $f_{\mathrm{sed}} \gtrsim 5$. Both of these planets exhibit decreased H$_2$O prominence for $m \gtrsim 10 \times$ solar, as the regions of maximal prominence shift to higher temperatures. Interestingly, Epsilon Eridani b thus represents a case where H$_2$O features $\lesssim$ 0.8 $\micron$ are more prominent for high $f_{\mathrm{sed}}$ and low $m$, whilst features around 0.94 $\micron$ become more prominent for low $f_{\mathrm{sed}}$ and high $m$. Given that $f_{\mathrm{sed}}$ and $m$ are not known for these planets, Epsilon Eridani b may thus represent a good target where simultaneous observations at both short and long wavelengths optimises the prospect of detecting H$_2$O. Guided by the picture offered by Figure \ref{fig:contour_g-T}, we proceed to examine self-consistent reflection spectra of HD 192310 c -- the planet with the overall highest predicted H$_2$O prominence -- in order to assess its potential for direct imaging spectroscopy.

\subsection{Assessment of HD 192310 c} \label{subsec:hc192310c}

\begin{figure*}[ht!]
\centering
\includegraphics[width=\textwidth]{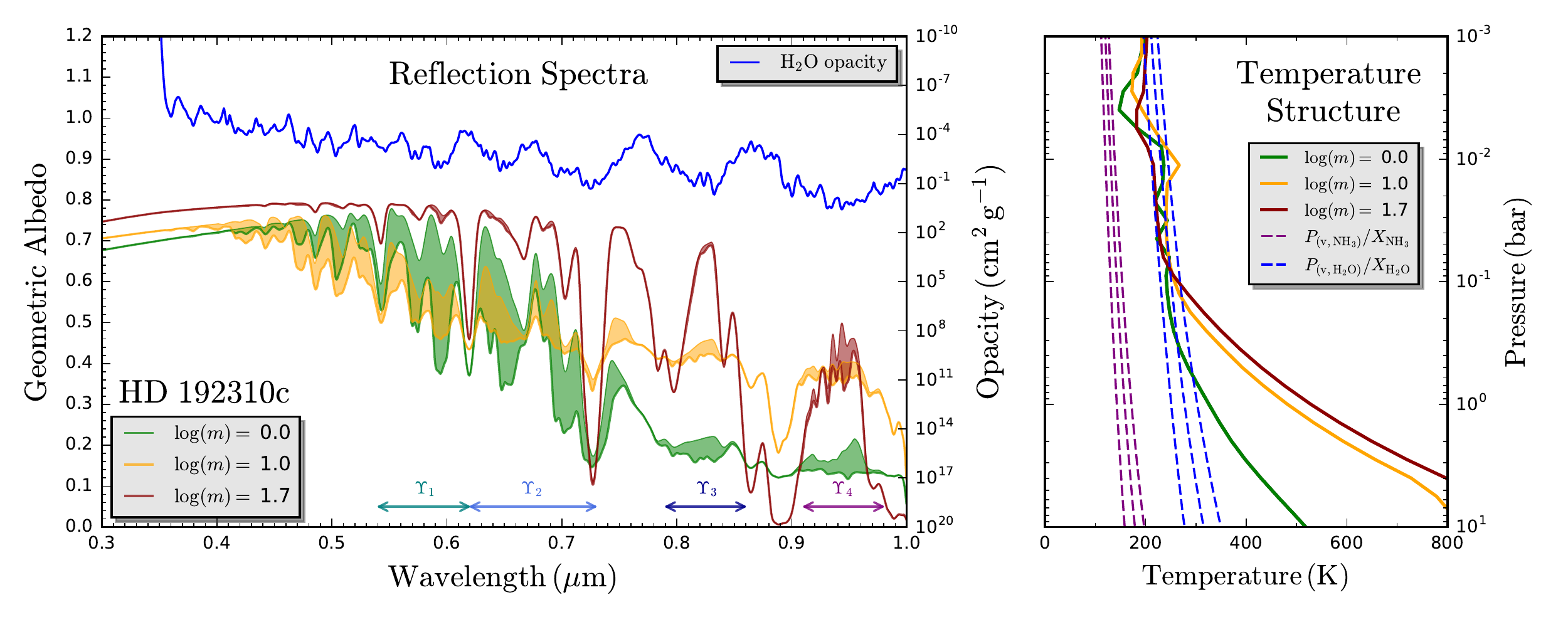}
\label{fig:HD192310c_spectra}
\caption{Self-consistent reflection spectra and P-T profiles for HD 192310c. Left: geometric albedo spectra for three models of HD 192310 c's atmosphere: $m = 1 \times$ solar (green), $m = 10 \times$ solar (orange), and $m = 50 \times$ solar (red). Two curves are plotted for each model; the lower curve has H$_2$O opacity enabled, the upper has H$_2$O opacity disabled, with the enclosed shading indicating absorption due to H$_2$O. All models assume $T_{\mathrm{eff}}$ = 190 K, $f_{\mathrm{sed}}$ = 3, and $g$ = 3.72 ms$^{-2}$. The H$_2$O opacity is shown in blue (smoothed for clarity). Right: P-T profiles corresponding to the same three models, with the NH$_3$ (purple) and H$_2$O (blue) condensation curves for each value of $m$ (increasing from left to right).}
\end{figure*}

The HD 192310 (Gliese 785) system, located 8.82 pc from Earth, contains two known Neptune-mass exoplanets with semi-major axes of $0.32 \pm 0.005$ AU and $1.18 \pm 0.025$ AU, respectively \citep{Pepe2011}. The outermost planet, HD 192310 c, has a slightly eccentric ($e = 0.32 \pm 0.11$) $526 \pm 9$ day orbit with an estimated equilibrium temperature of 185 K. Radial velocity observations with the HARPS instrument have measured $M_{p} \, \mathrm{sin}(i) = 24 \pm 5$ $M_{\Earth}$ ($0.076 \pm 0.016 \, M_{J}$) \citep{Pepe2011}. These system parameters place HD 192310 c near the currently proposed inner working angle of WFIRST \citep{Nayak2017}, making direct imaging observations of this planet challenging but not impossible.

In order to model geometric albedo spectra of this planet, some assumptions must be made. Given the unknown radius, we use the mass-radii relations of \citet{Fortney2007} to infer $R_{p} \approx 0.71 R_{J}$. Taking $M_p \, \mathrm{sin}(i) = M_p$, we derive a surface gravity of $g$ = 3.72 ms$^{-2}$. Using a planetary evolution model \citep{Fortney2008a}, we find $T_{\mathrm{eff}} = 190_{-25}^{+40}$ K. A reasonable value of $f_{\mathrm{sed}}$ = 3 is chosen. Unlike our models in previous sections, here we use P-T profiles that self-consistently include the effect of clouds via coupling to the \citet{Ackerman2001} cloud model. Atmospheres with five different metallicities are considered: $\mathrm{log} (m)$ = 0.0, 0.5, 1.0, 1.5, and 1.7 $\times$ solar (corresponding to $m$ = 1, 3.16, 10, 31.6, and 50 $\times$ solar). The resulting geometric albedo spectra, previously shown in \citet{Nayak2017}, and P-T profiles for the solar, $10 \times$ solar, and $50 \times$ solar models are presented in Figure \ref{fig:HD192310c_spectra}.

Our self-consistent models of HD 192310 c confirm that H$_2$O absorption can play a prominent role in shaping its geometric albedo spectrum. In particular, the prediction from Sections \ref{sec:parameters} and \ref{sec:max_detectability} that planets with $g \lesssim$ 20 ms$^{-2}$ should display significant H$_2$O signatures at wavelengths $\lesssim 0.8 \, \micron$ is vindicated. We also see that the predicted trend of $m \gtrsim 10 \times$ solar brightening the continuum albedo, but resulting in severely dampened short wavelength H$_2$O signatures (e.g. Figure \ref{fig:vary_m}), holds for the self-consistent models. We note that the self-consistent models do exhibit one additional feature, namely that the presence of multiple cloud decks can be caused by high-altitude thermal inversions. Such an effect was not encountered in our parameter space exploration, as our parametric P-T profile (Section \ref{subsubsec:PT_profile}) has a temperature structure that monotonically decreases with altitude. However, this does not overly modify our earlier conclusions, as these high-altitude cloud decks tend to be optically thin.

The prominence of H$_2$O absorption and key cloud properties for each of our HD 192310 c atmosphere models are summarised in Table \ref{table:HD192310c_results}. H$_2$O prominence is expressed in terms of the spectral indices $\Upsilon_{1,2,3,4}$ (Section \ref{subsec:quantify_indices}), with clouds summarised by the cloud base pressure, $P_{\mathrm{base}}$, cloud deck thickness, $\Delta P_{\mathrm{cld}}$, and the maximum layer optical depth within each cloud deck, $\tau_{\mathrm{cld, \, max}}$. The asymmetry parameter and single scattering albedo are essentially the same for all H$_2$O clouds ($\bar{g} \approx 0.89$, $\bar{\omega} \approx 0.999$). Comparing with Figure \ref{fig:contour_g-T}, our parameter space survey predicted $\Upsilon_{1} \approx 0.22$ for $\mathrm{log} \, (m) = 0.5 \times$ solar, which matches well with the self-consistent value of 0.173. Our predicted value of $\Upsilon_{4} \approx 0.04$ agrees with the higher metallicity models, but under-predicts the prominence of H$_2$O at 0.94 $\micron$ yielded by the self-consistent model (0.123). This is caused by brightening by the sudden transition to an optically thick H$_2$O cloud deck when $m$ first increases to $\sim 3 \times$ solar -- due to the H$_2$O condensation curve shifting to higher temperatures, resulting in a deeper intersection point.

\begin{deluxetable}{cccccccc}
%\tablewidth{100pt}
\tablecaption{Theoretical H$_2$O Prominence and Cloud Properties for HD 192310 c}
\tablehead{
\colhead{$\mathrm{log} \, (m)$} & \colhead{$\Upsilon_{1}$ \tablenotemark{a}} & \colhead{$\Upsilon_{2}$} & \colhead{$\Upsilon_{3}$} & \colhead{$\Upsilon_{4}$} & \colhead{$P_{\mathrm{base}}$} & \colhead{$\Delta P_{\mathrm{cld}}$ \tablenotemark{b}} & \colhead{$\tau_{\mathrm{cld, \, max}}$} \\ 
\colhead{($\times$ solar)} & \colhead{} & \colhead{} & \colhead{} & \colhead{} & \colhead{(mbar)} & \colhead{(mbar)} & \colhead{} } 
\startdata
0.0 & 0.300 & 0.227 & 0.020 & 0.019 & 16 & 12 & 0.7 \\
0.5 & 0.173 & 0.137 & 0.037 & 0.123 & 4 / 251 \tablenotemark{c} & 2 / 162 & 0.7 / 13 \\
1.0 & 0.145 & 0.097 & 0.025 & 0.035 & 8 / 126 & 5 / 63 & 5.2 / 3.6 \\
1.5 & 0.011 & 0.018 & 0.016 & 0.054 & 8 / 251 & 4 / 206 & 0.6 / 239 \\
1.7 & 0.009 & 0.014 & 0.009 & 0.040 & 11 / 251 & 5 / 229 & 0.7 / 304 \\
\enddata
\tablenotetext{a}{Spectral indices $\Upsilon{i}$, as defined by Equation \ref{eq:spec_index_equation}.}
\tablenotetext{b}{Thickness of cloud deck(s) above cloud base.}
\tablenotetext{c}{Slashes (/) indicate multiple cloud decks.}
\vspace{-1.0cm}
\label{table:HD192310c_results}
\end{deluxetable}

The models with $\mathrm{log} \, m \geq 0.5 \times$ solar also exhibit secondary high altitude cloud decks at $\sim 1-10$ mbar. These are are a consequence of weak thermal inversions in the upper atmosphere, leading to multiple crossings of the relevant H$_2$O condensation curves. This does not occur for the solar metallicity model, which only intersects the condensation curve once at 16 mbar. These secondary cloud decks are optically thin in most cases, with the one exception being our $\mathrm{log} \, m = 1.0 \times$ solar model, for which the two cloud decks have similar optical depths. As shown in Figure \ref{fig:HD192310c_spectra}, two distinct, moderately scattering, cloud decks still yield especially prominent H$_2$O features for $\lambda \lesssim 0.7 \, \micron$. These H$_2$O signatures in particular underscore the importance of low $g$ planets to observations.

\section{Summary And Discussion} \label{sec:discussion}

We have conducted an extensive parameter space survey to assess the prominence of H$_2$O absorption signatures in reflection spectra of cool giant exoplanets. We have quantitatively explored the impact of effective temperature, gravity, sedimentation efficiency, and metallicity, providing $>$ 50,000 reflection spectra models for the community\footnote{\href{https://doi.org/10.5281/zenodo.1210305}{Reflection Spectra Repository}}. Our main conclusions are as follows: 

\begin{itemize}
   \item Giant planets $\gtrsim 30$ K warmer than Jupiter can exhibit reflection spectra substantially altered by H$_2$O.
    \item Prominent H$_2$O absorption features exist from $\sim 0.91 - 0.98 \, \micron$, $\sim 0.79 - 0.86 \, \micron$, and are embedded in the Rayleigh slope from $\sim 0.4 - 0.73 \, \micron$.
    \item Planets with $g \lesssim$ 20 ms$^{-2}$ and $T_{\mathrm{eff}} \gtrsim 180$ K exhibit the most prominent H$_2$O features, manifestly observable at short wavelengths $\lesssim 0.7 \, \micron$. This holds across a wide range of possible values for $m$ and $f_{\mathrm{sed}}$.
	\item Planets with $g \gtrsim$ 20 ms$^{-2}$ and $T_{\mathrm{eff}} \sim 150$ K can exhibit notable H$_2$O features around 0.94 $\micron$, given $f_{\mathrm{sed}} \gtrsim 3$ and $m \lesssim 10 \times$ solar.
	\item Higher $f_{\mathrm{sed}}$ enables lower temperature planets to support prominent H$_2$O features.
	\item Higher $m$ tends to weaken H$_2$O features, resulting in bright planets dominated by scattering H$_2$O clouds.
    \item HD 192310 c is identified as the most promising currently known cool giant planet to detect H$_2$O via direct imaging in reflected light.
\end{itemize}

Of all our results, we draw particular attention to the finding that low gravity planets possess exceptionally strong H$_2$O features along the Rayleigh slope from $\sim 0.4 - 0.73 \, \micron$. To our knowledge, observable H$_2$O features at such short wavelengths have not previously been discussed in the literature, so our finding that they are often the most potent H$_2$O signatures bears note. The enhanced presence of such features for low gravity planets is due to the combination of optically thin clouds and large density scale heights -- resulting in long path lengths for relatively cloud-free atmospheres. On the other hand, the 0.94 $\micron$ H$_2$O feature cannot rely on a Rayleigh scattering continuum, and so is less prominent despite possessing a H$_2$O opacity exceeding that at shorter wavelengths by orders of magnitude. Indeed, the 0.94 $\micron$ feature critically depends on clouds to brighten the albedo at these wavelengths, which results in narrow regions of parameter space where it may be observed. Conversely, the $\sim 0.4 - 0.73 \, \micron$ H$_2$O features need only require a relatively cloud-free low gravity planet to be prominent, which holds true over a vast expanse of parameter space.

We have explicitly highlighted, via self-consistent modelling of HD 192310 c, the substantial imprint of H$_2$O in a low gravity planet's reflection spectrum. As this is the only known radial velocity planet amenable to direct imaging with an estimated $g \lesssim$ 20 ms$^{-2}$, there is an urgent need to identify additional Neptune-mass exoplanets with $T_{\mathrm{eff}} \gtrsim 180$ K to capitalise on `gravity enhanced' H$_2$O signatures. 

Our results demonstrate a profound advantage for spectroscopic observations at short wavelengths from $\sim 0.4 - 0.73 \, \micron$. WFIRST spectroscopy, by comparison, is envisioned to have a minimum observable wavelength of 0.6 $\micron$ \citep{Spergel2015}, resulting in a narrow range where gravity enhanced H$_2$O signatures can be harnessed. Future direct imaging missions, such as LUVOIR or HabEx, should aim for the greatest flexibility to obtain spectra over large ranges of $\lambda / D$, where $D$ is telescope diameter, to maintain the ability to fully exploit this effect for a large variety of planets.

Interpretations of direct imaging observations will additionally have to contend with effects not considered here. Observational constraints will necessitate imaging at larger phase angles than the zero phase angle geometric albedo spectra we have examined -- typically between 40 - 130$\degr$ for edge-on systems with WFIRST \citep{Mayorga2016}. Under the typical observing condition of quarter-phase, partial illumination can result in reflection spectra darker by a factor of 3-4 around 0.55 $\micron$ and 0.75 $\micron$ \citep{Sudarsky2005,Mayorga2016}. Inferences of H$_2$O features $\lesssim 0.5 \, \micron$ will additionally require modelling of photochemical hazes, which can substantially darken short wavelength albedo spectra \citep{Marley1999,Sudarsky2000,Gao2017}. Furthermore, our modelling framework explicitly assumes abundances of non-condensing species are governed by thermochemical equilibrium. The atmospheres of both Jupiter and Saturn contain disequilibrium gasses (e.g., $\rm PH_3$) delivered to the observable atmosphere by vertical mixing \citep{Atreya1999}, whilst evidence of disequilibrium chemistry from exoplanet atmospheric retrievals has recently emerged from studies of exo-Neptunes \citep{Morley2017}, directly imaged planets \citep{Lavie2017}, and hot Jupiters \citep{MacDonald2017b}. We thus should not be surprised if the atmospheres of cool giant exoplanets are not in chemical equilibrium.

Beyond detecting H$_2$O, our results suggest that it may be possible to infer the mixing ratio of H$_2$O from atmospheric retrievals of reflection spectra. The prominence of H$_2$O signatures can rival that of CH$_4$, particularly for low gravity planets, and so abundance constraints could be derived from low-resolution observations similar to that expected of WFIRST ($R \approx 50$). Atmospheric retrieval techniques for reflection spectra have seen development in recent years \citep{Barstow2014,Lupu2016,Nayak2017,Lacy2018}, but much of this work has focused on CH$_4$ as the dominant molecular opacity source. Our present study demonstrates the need to include H$_2$O vapour as a parameter in reflection spectra retrievals. Indeed, the sensitivity of H$_2$O features to the value of $g$ offers a potential solution to the issue whereby $g$ is poorly constrained when retrieving CH$_4$ alone \citep{Lupu2016}. Ultimately, the application of such retrieval methods offers the prospect of deriving constraints on the C/O, O/H, and other elemental ratios for cool giant planets. Such elemental ratios, constrained across a wide sample of extrasolar Jupiter analogues, offers extraordinary promise for resolving outstanding questions on the formation mechanisms for cool giant planets.

\hfill

The opportunity to conduct this research was enabled by the 2016 Kavli Summer Program in Astrophysics, supported by grants from The Kavli Foundation, The National Science Foundation, UC Santa Cruz, and UCSC's Other Worlds Laboratory. In particular, we commend P. Garaud, the program coordinator, and the Scientific Organising Committe (J. Fortney, D. Abbot, C. Goldblatt, R. Murray-Clay, D. Lin, A. Showman and X. Zhang). R.J.M. additionally acknowledges financial support from the Science and Technology Facilities Council (STFC), UK, toward his doctoral program. We thank P. Gao for insightful discussions on geometric albedo spectra.

\appendix

\setcounter{figure}{0}
\setcounter{table}{0}
\renewcommand{\thefigure}{\Alph{figure}}
\renewcommand{\thetable}{\Alph{table}}

\section{A Pressure-Temperature Profile for Cool Giant Planets} \label{Appendix}

Here we describe the origin of the P-T profiles employed in our cool giant planet model grid. Our aim was to develop a simple empirical fitting function capable of approximately reproducing the temperature structure of cool giant planets under radiative-convective equilibrium, thereby circumventing the computational cost of running thousands of equilibrium models directly.

We began with a set of 68 self-consistent P-T profile models computed using the methods described in \citet{Fortney2008}. These profiles model giant planets around a solar-analogue with $g$ from 10 ms$^{-2}$ to 100 ms$^{-2}$, $m$ from 1$\times$ to 100$\times$ solar, and $a$ from 0.5 AU to 5.0 AU. All models have $T_{\mathrm{int}}$ = 150 K, C/O = 0.5, and assume a cloud-free atmosphere under radiative-convective equilibrium. The model atmospheres are divided into 60 layers, specified at 61 depth points uniformly spaced in log-P from $10^{-6}$ bar to $10^{3}$ bar. $T_{\mathrm{eff}}$ was derived for each model from integrating the thermal emission. Each model is thus specified by $g$, $\mathrm{log} \, (m)$, and $a$ (or $T_{\mathrm{eff}}$). The parameters describing each of the models used are given in Table \ref{table:fortney_models}.

For each of these self-consistent P-T profile models, we fit a simple two-parameter function:

\begin{equation}
T^{4} (P) = T_{0}^{4} + T_{\mathrm{deep}}^{4} (P/1000 \mathrm{bar}) \tag{4}
\end{equation}
where $T$ is the atmospheric temperature at pressure $P$ (in bar), and the free parameters are $T_{0}$ and $T_{\mathrm{deep}}$. This functional form is essentially just the Eddington relation with negligible irradiation ($T_{\mathrm{irr}} \ll T_{\mathrm{int}}$), with the grey optical depth, $\tau$, taken to be proportional to $P$ and with the various constants absorbed into the parameters $T_{0}$ and $T_{\mathrm{deep}}$. Physically, $T_{0}$ and $T_{\mathrm{deep}}$ encode the temperatures of an upper-atmosphere isotherm and the temperature at the 1000 bar level.

We derived best-fitting values of the parameters $T_{0}$ and $T_{\mathrm{deep}}$ using a technique borrowed from exoplanet atmospheric retrieval. For a given self-consistent model, the temperature in each layer, $T_{\mathrm{Fortney}} \, (P_i)$, can be taken as a `data-point'. For a given point in the $T_{0}$-$T_{\mathrm{deep}}$ plane, a `model' temperature profile, $T_{\mathrm{fit}}\, (P_i)$, can be constructed using Equation \ref{eq:PT_parametrisation}. By comparing the `data' to the `model', a likelihood function can be constructed:

\begin{equation}
\mathrm{log} \mathcal{L} = \sum_{i = 1}^{N} - \frac{[T_{\mathrm{Fortney}} \, (P_i) - T_{\mathrm{fit}} (T_{0}, T_{\mathrm{deep}}, P_i)]^{2}}{2 \delta_{T}^{2}}  \tag{A.1}
\label{eq:likelihood}
\end{equation}
where $N$ is the number of depth points (61), and $\delta_{T}$ is an arbitrary `temperature tolerance' factor that encodes an `error' in the self-consistent model temperatures. This functional form essentially assumes independently distributed Gaussian `errors', with an additive normalisation factor in the log-likelihood neglected (due to it being identical for all sampled values of $T_{0}$ and $T_{\mathrm{deep}}$). For our purposes, where we essentially assume the self-consistent models represent `ground truth', we take $\delta_{T}$ = 1 K (smaller values increase the fitting time, with negligible influence on the final derived fits). We employ the MultiNest algorithm \citep{Feroz2008,Feroz2009} to map the $T_{0}$ and $T_{\mathrm{deep}}$ plane, bounded by wide uniform priors from 0 K to 1000 K for $T_{0}$ and 600 K to 3000 K for $T_{\mathrm{deep}}$. We ran MultiNest once for each self-consistent profile (typically involving $\sim 10^5$ likelihood evaluations), drawing the maximum likelihood posterior sample values of $T_{0}$ and $T_{\mathrm{deep}}$ to arrive at the optimal fitted profile.

\begin{figure*}[t!]
\centering
\includegraphics[width=0.99\textwidth]{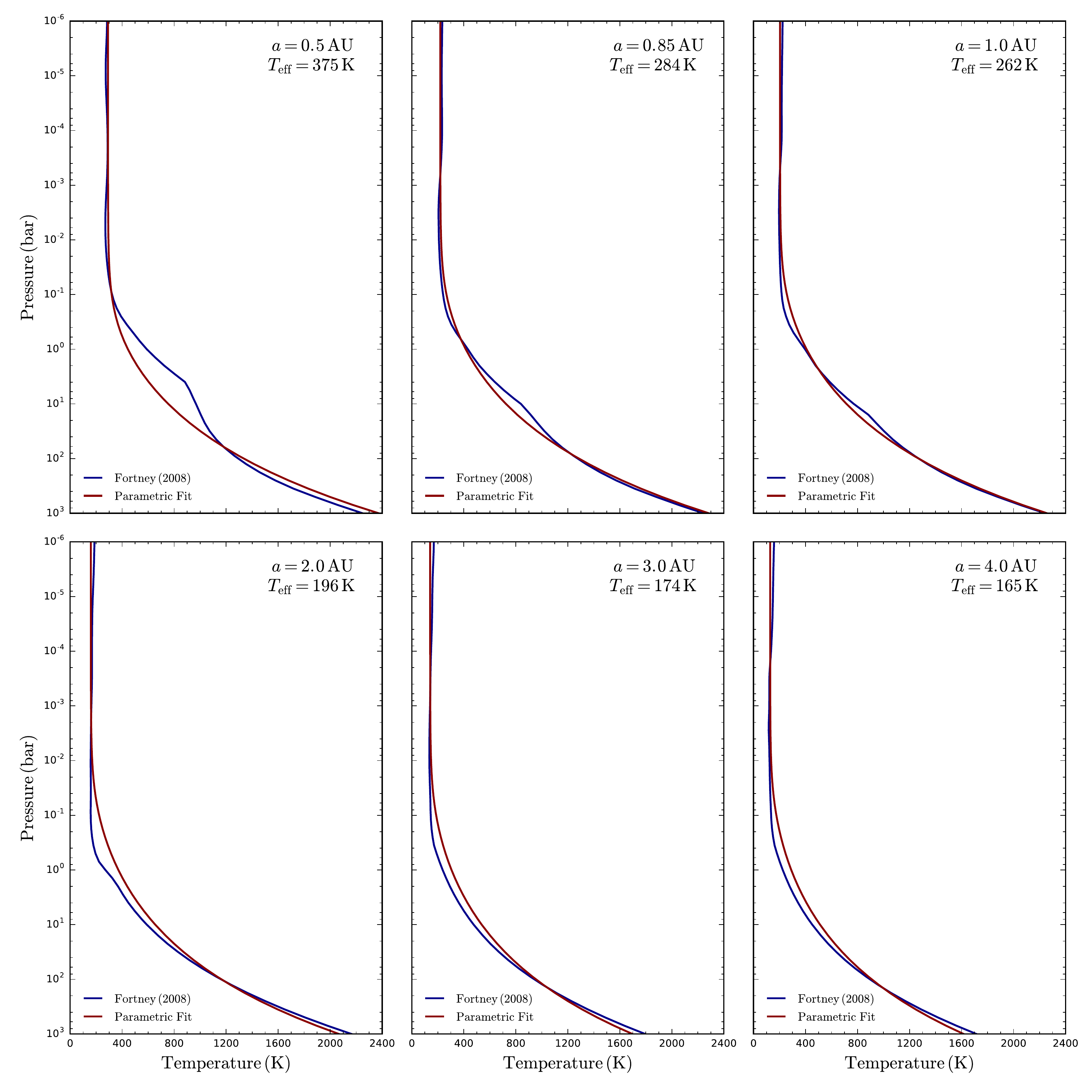}
\label{fig:PT_fits}
\caption{Comparison between self-consistent cool giant planet P-T profiles \citep{Fortney2008} (blue) and the best-fitting P-T profiles described by Equation \ref{eq:PT_parametrisation} (red). The six profiles are for a solar metallicity cloud-free Jovian analogue ($g$ = 25 ms$^{-2}$) with $T_{\mathrm{int}}$ = 150 K, and C/O = 0.5, with planet-star separations from 0.5 to 4.0 AU. The values of $T_{\mathrm{eff}}$ corresponding to each model are given in the top right of each panel.}
\end{figure*}

We compare the fitted profile using Equation \ref{eq:PT_parametrisation} to the self-consistent input profiles for six models in Figure \ref{fig:PT_fits}. The plotted models are for a solar metallicity Jupiter analogue ($g$ = 25 ms$^{-2}$) with semi-major axes from 0.5 AU to 4.0 AU. The five models with $T_{\mathrm{eff}} < 300$ K are generally well-fit, with maximum deviations between the self-consistent P-T profile and the fitted profile $<$ 120 K (with mean deviations $<$ 40 K). Discrepancies grow for models with $T_{\mathrm{eff}} > 300$ K, as shown in the upper left panel of Figure \ref{fig:PT_fits}, due to the parametrisation of Equation \ref{eq:PT_parametrisation} being unable to account for the influence of non-negligible external radiation. For this case ($a = 0.5$ AU), the maximum deviation between the self-consistent P-T profile and the fitted profile is 280 K (with a mean deviations of 60 K). We thus note that the maximum effective temperature over which our cool giant planet P-T profile fits should be utilised is $\sim$ 300 K (it is for this reason that Figures \ref{fig:contour_f-T} and \ref{fig:contour_g-T} do not go above 300 K). Similarly, our profile should not be used for $T_{\mathrm{eff}} < 150$ K, as $T_{\mathrm{eff}}$ cannot be less than $T_{\mathrm{int}}$ (150 K for all the input self-consistent profiles).

\begin{figure*}[ht!]
\centering
\includegraphics[width=\textwidth]{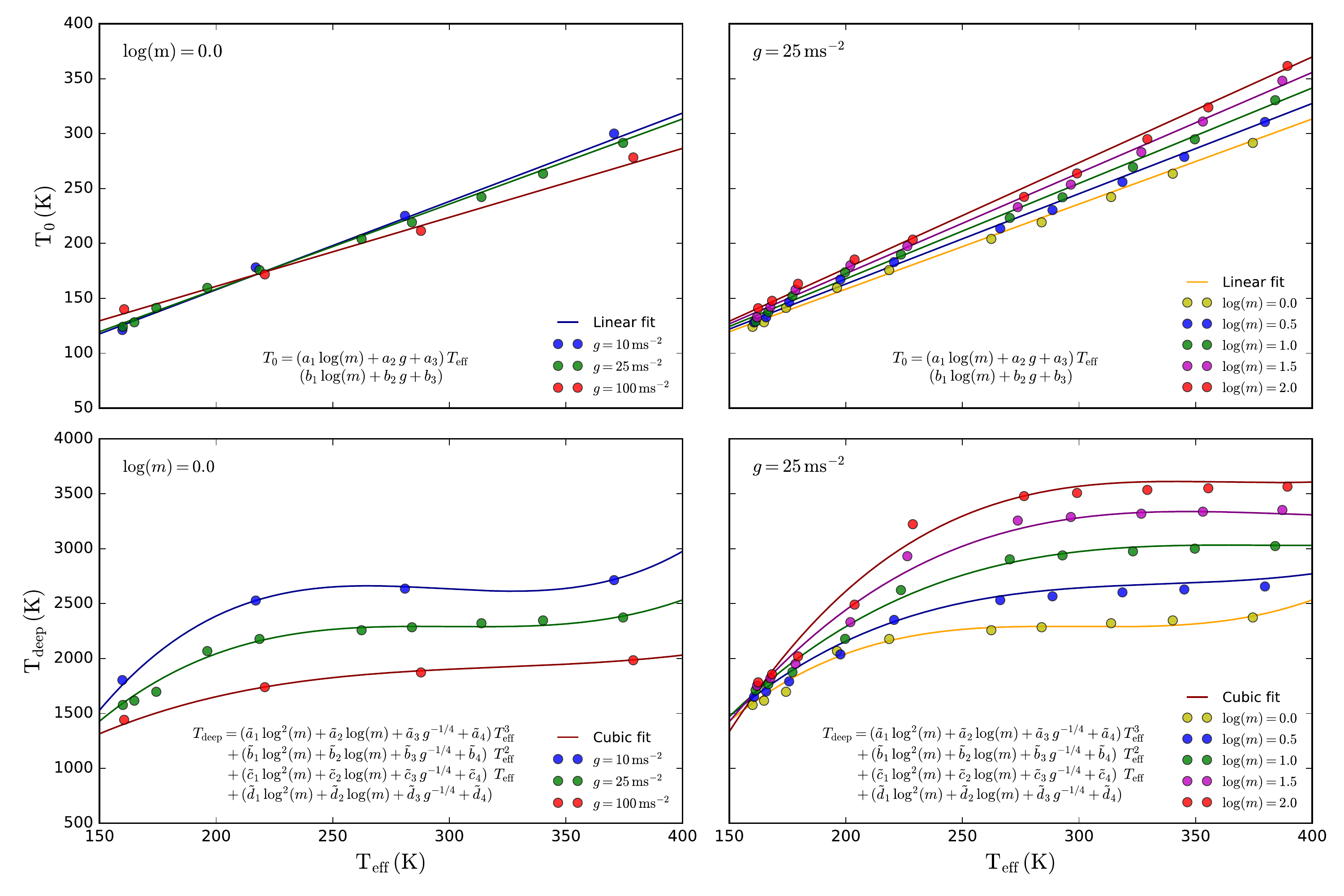}
\label{fig:PT_param_fits}
\caption{Variation of the best-fitting values of $T_{0}$ and $T_{\mathrm{deep}}$ with metallicity, gravity, and effective temperature. Left: variation with $g$ and $T_{\mathrm{eff}}$ for fixed $m$. Right: variation with $m$ and $T_{\mathrm{eff}}$ for fixed $g$. The points show the values of $T_{0}$ and $T_{\mathrm{deep}}$ that maximise Equation \ref{eq:likelihood} for each of the self-consistent models in Table \ref{table:fortney_models}. The curves are fitting functions to the points, expressed in Equations \ref{eq:T0} and \ref{eq:Tdeep} (numerical values of the expansion coefficients are given in the text).}
\end{figure*}

The best-fitting values of $T_{0}$ and $T_{\mathrm{deep}}$ for each of the 68 self-consistent P-T profiles are plotted as functions of $T_{\mathrm{eff}}$ in Figure \ref{fig:PT_param_fits}. We investigated a number of functional forms for $T_{0} (\mathrm{log}(m), \, g, \, T_{\mathrm{eff}})$ and $T_{\mathrm{deep}} (\mathrm{log}(m), \, g, \, T_{\mathrm{eff}})$ able to reproduce the broad trends in these quantities across parameter space. Least-squares fitting for each functional form was conducted for fixed $m$ to derive fitting coefficients for variations in $g$ and for fixed $g$ to derive coefficients for variations in $m$. Coefficients were matched at the common intersection point of these two families of solutions ($g$ = 25 ms$^{-2}$ and $\mathrm{log}(m)$ = 0.0).

For $T_{0}$, a linear relation with $T_{\mathrm{eff}}$ was found to be adequate, taking the form:

\begin{equation}
T_0 = (a_1 \, \mathrm{log}(m) + a_2 \, g + a_3) \, T_{\mathrm{eff}} + 
(b_1 \, \mathrm{log}(m) + b_2 \, g + b_3) \tag{A.2}
\label{eq:T0}
\end{equation}
where $a_1 = 0.0939629520975$, $a_2 = -0.00195955723764$, $a_3 = 0.824004813287$, $b_1 = -9.3354717917$, $b_2 = 0.426544316687$, and $b_3 = -7.35793593873$. For $T_{\mathrm{deep}}$, the minimal order polynomial fitting function was found to be cubic in $T_{\mathrm{eff}}$:

\begin{equation}
\begin{aligned}
T_{\mathrm{deep}} =  & (\tilde{a}_1 \, \mathrm{log}^2(m) + \tilde{a}_2 \, \mathrm{log}(m) + \tilde{a}_3 \, g^{-1/4} + \tilde{a}_4) \, T_{\mathrm{eff}}^{3} +
(\tilde{b}_1 \, \mathrm{log}^2(m) + \tilde{b}_2 \, \mathrm{log}(m) + \tilde{b}_3 \, g^{-1/4} + \tilde{b}_4) \, \, \, T_{\mathrm{eff}}^{2} + \\
& (\tilde{c}_1 \, \mathrm{log}^2(m) + \tilde{c}_2 \, \mathrm{log}(m) + \tilde{c}_3 \, g^{-1/4} + \tilde{c}_4) \, \, \, T_{\mathrm{eff}} +
(\tilde{d}_1 \, \mathrm{log}^2(m) + \tilde{d}_2 \, \mathrm{log}(m) + \tilde{d}_3 \, g^{-1/4} + \tilde{d}_4) 
\end{aligned}
\label{eq:Tdeep} \tag{A.3}
\end{equation}
where $\tilde{a}_1$ = 0.000122391650778, $\tilde{a}_2$ = -0.000253005417439, $\tilde{a}_3$ = 0.00130682904558, $\tilde{a}_4$ = -0.000324145003492, $\tilde{b}_1$ = -0.101884473342, $\tilde{b}_2$ = 0.188364010495, $\tilde{b}_3$ = -1.1319726177, $\tilde{b}_4$ = 0.271721883808, $\tilde{c}_1$ = 26.8759214447, $\tilde{c}_2$ = -40.9475604372, $\tilde{c}_3$ = 317.541624812, $\tilde{c}_4$ = -71.729686904, $\tilde{d}_1$ = -2241.35380397, $\tilde{d}_2$ = 2888.7627511, $\tilde{d}_3$ = -25698.9658072, $\tilde{d}_4$ = 6778.41758375.

For given values of $\mathrm{log}(m)$, $g$, and $T_{\mathrm{eff}}$, substitution into Equations \ref{eq:T0} and \ref{eq:Tdeep} yield $T_0$ and $T_{\mathrm{deep}}$. Substituting these into Equation \ref{eq:PT_parametrisation} then yields an a priori determined P-T profile valid for cool giant planets with $T_{\mathrm{eff}}$ in the range $\sim$ 150-300 K.

\begin{deluxetable}{lll}
\tablecaption{\citet{Fortney2008} P-T models employed during parametric fitting of Equation \ref{eq:PT_parametrisation}.}
\tablehead{
\colhead{$g$} & \colhead{$\mathrm{log} \, (m)$} & \colhead{$a$} \\ 
\colhead{(ms$^{-2}$)} & \colhead{($\times$ solar)} & \colhead{(AU)} }
\startdata
10 & 0.0 & 0.5, 0.85, 1.5, 5.0 \\
25 & 0.0, 0.5, 1.0 & 0.5, 0.6, 0.7, 0.85, 1.0 \\
   & 1.5, 1.7, 2.0 & 1.5, 2.0, 3.0, 4.0, 5.0 \\
100 & 0.0 & 0.5, 0.85, 1.5, 5.0 \\
\enddata
\tablenotetext{}{}
\label{table:fortney_models}
\end{deluxetable}

%\bibliographystyle{aasjournal}
%\bibliography{reflected_light}

\listofchanges

\end{document}